\documentclass[a4paper,fleqn,usenatbib]{mnras}

\usepackage{txfonts}

\usepackage[T1]{fontenc}
\usepackage{ae,aecompl}

\usepackage{graphicx}	
\usepackage{braket}
\usepackage{cprotect}

\def\mr#1{\mathrm{#1}}

\title[The cold front in Abell~3667]{An azimuthally resolved study of the cold front in Abell~3667}

\author[Y. Ichinohe et al.]{
Y. Ichinohe,$^{1}$\thanks{E-mail: ichinohe@tmu.ac.jp}
A. Simionescu,$^{2}$
N. Werner$^{3,4}$
and T. Takahashi$^{2}$
\\
$^{1}$Department of Physics, Tokyo Metropolitan University, 1-1 Minami-Osawa, Hachioji, Tokyo 192-0397, Japan\\
$^{2}$Institute of Space and Astronautical Science (ISAS), JAXA, 3-1-1 Yoshinodai, Chuo, Sagamihara, Kanagawa 252-5210, Japan\\
$^{3}$MTA-E\"{o}tv\"{o}s University Lend\"{u}let Hot Universe Research Group, P\'{a}zm\'{a}ny P\'{e}ter s\'{e}t\'{a}ny 1/A, Budapest, 1117, Hungary\\
$^{4}$Department of Theoretical Physics and Astrophysics, Faculty of Science, Masaryk University, Kotl\'a\v{r}sk\'a 2, Brno, 611 37, Czech Republic
}

\date{}

\pubyear{2016}

\begin{document}
\label{firstpage}
\pagerange{\pageref{firstpage}--\pageref{lastpage}}
\maketitle

\begin{abstract}
 The microphysical properties, such as effective viscosity and conductivity, of the weakly magnetized intergalactic plasma are not yet well known. We investigate the constraints that can be placed by an azimuthally resolved study of the cold front in Abell~3667 using $\sim$500~ksec archival {\it Chandra} data. We find that the radius of the interface fluctuates with position angle and the morphology of the interface is strikingly similar to recent numerical simulations of inviscid gas-stripping. We find multiple edges in the surface brightness profiles across the cold front as well as azimuthal variations, which are consistent with the presence of Kelvin-Helmholtz Instabilities (KHI) developing along the cold front. They indicate that the characteristic length scale of KHI rolls is around 20--80~kpc. This is the first observational indication of developing KHIs along a merger cold front in a galaxy cluster. Based on the KHI scenario, we estimated the upper limit of the ICM effective viscosity. The estimated value of $\mu\lesssim200~\mr{g/cm/s}$ is at most 5\% of the isotropic Spitzer-like viscosity. The observed apparent mixing towards the outer edges away from the tip of the front provides an additional evidence for suppressed viscosity.
\end{abstract}




\section{Introduction}
 Among gravitationally collapsed structures, clusters of galaxies are the largest and most recently formed (and still forming). They evolve to the typical mass of 10$^{14\mathchar`-15}M_\odot$ via accretion of smaller structures and successive mergers of smaller clusters or groups \citep{sarazin02}. Most of the intergalactic medium (IGM) currently observable resides in galaxy clusters where it is referred to as intracluster medium (ICM).

Although most of the baryons reside in the IGM/ICM, many questions remain about the physics of this diffuse plasma. The most fundamental and important open issues regard the basic microphysical properties of the IGM, such as its effective viscosity, thermal conductivity, geometry and strength of the magnetic fields, turbulence, mixing time-scale, and heating-cooling balance.

Abell~3667 is a nearby \citep[$z=0.055$,][]{sodre92} non-cool-core cluster of galaxies known for the various indications of its recent merger in a wide range of electromagnetic wavelengths: radio, optical and X-ray. Early optical and X-ray observations of the system revealed elongated double-peaked X-ray morphology in the northwest-southeast direction, in which the locations of the two brightest galaxies coincide with the two X-ray peaks \citep{sodre92}. A more in depth optical study by \citet{owers09a} showed that the member galaxy distribution is significantly bimodal, and the offset of the peculiar velocities of the two galaxy distributions is $\sim$500$~\mr{km/s}$.

In the radio band, its most prominent feature is the northwestern extended radio emission \citep{rottgering97}. This feature is classified as a radio relic, and is the brightest known source in this class. The radio spectral index shows steepening from the outer edge toward the cluster centre, indicating the aging of the non-thermal electrons \citep{hindson14}. At the opposite side of the radio relic with respect to the cluster centre, less prominent radio relics are also observed \citep{rottgering97}. \citet{roettiger99} studied the system numerically and indicated that these relics are one of the consequences of merger activity. In addition to the relics, it has been suggested recently that Abell~3667 also hosts another type of diffuse radio emission, which connects these relics (radio bridge), and which may be associated with the ICM turbulence due to this merger \citep{carretti13,riseley15}. The overall radio structure is aligned well with the major axis of the X-ray emission and the axis of the galaxy bimodality.

The northwestern radio relic is extensively studied also in the X-ray band. \citet{finoguenov10} and \citet{sarazin13} investigated the thermodynamic structure around the relic using {\it XMM-Newton}, and observed a sharp drop of the temperature and the surface brightness, indicating that the origin of the relic is a merger shock with the Mach number of $M\sim 2$. Subsequent {\it Suzaku} observations also confirmed the result \citep{akamatsu12a,akamatsu13}, further indicating that the plasma may be out of thermal equilibrium in this region. Its non-thermal nature has also been extensively explored, but the problem of whether the X-ray emission associated with the relic is thermal or non-thermal is still to be resolved \citep{fuscofemiao01,rephaeli04,nakazawa09,finoguenov10,akamatsu12a}. 

Initially, the X-ray brightness edge opposite to the northwestern relic was thought to be also a shock front \citep{markevitch99}. The {\it Chandra} observation of the brightness drop surprisingly revealed that this is not the case, but instead, the edge is a ``cold front'' which is the interface between a cool, dense gas volume and the hot, tenuous ambient medium \citep{vikhlinin01b}. Abell~3667 is one of the first clusters in which a cold front was observed while, to date, it is known that cold fronts are observationally more common than shock fronts \citep{markevitch07,owers09b}.

The thinness of the cold front, which cannot be resolved even with the {\it Chandra} angular resolution, has offered us a wealth of indications regarding ICM microphysics. \citet{vikhlinin01b} first suggested that transport processes are heavily suppressed across the interface, and also pointed out the absence of hydrodynamic instabilities at the front. \citet{vikhlinin01a} estimated the magnetic field strength required to keep the front hydrodynamically stable. \citet{vikhlinin02} considered the role of gravity for the stability, while \citet{churazov04b} suggested the intrinsic width of the interface can also stabilize the front. \citet{mazzotta02} pointed out a signature of a possible very-large-scale developing instability.

The cold front is also studied numerically by e.g., \citet[][]{heinz03}, and the predictions are confirmed by deeper data: the thermodynamic maps based on the {\it XMM-Newton} and {\it Chandra} observations revealed that cold, low-entropy and high-metallicity gas is uplifted toward the tip of the front \citep{briel04,lovisari09,datta14}, while the shape of the front itself was observed to be mushroom-like \citep{owers09b}.

Because of its prominence and proximity, this cold front is a rather well studied one, as mentioned above. However, the studies done so far have not focused on the azimuthal information of the front. Since hydrodynamic instabilities occur on the interface, the azimuthal variation should contain a lot of information about the microphysical properties of the ICM, in a similar way as the radial dependence does. Abell~3667 is relatively nearby, and the front is very prominent, with a large opening angle, which make it the ideal target for such a study regarding the azimuthal variation. This is the topic of the present manuscript.

 Unless otherwise noted, the error bars correspond to 68\% confidence level for one parameter. Throughout this paper, we assume the standard $\Lambda$CDM cosmological model with the parameters of $(\Omega_m,\Omega_\Lambda,H_0)=(0.3,0.7,70~\mr{km/s/Mpc})$.

\section{Observations, data reduction, and data analysis}\label{sec:data}

\begin{table}
 \centering
 \caption{Summary of the observations used in this paper. The net exposure time is after the data screening.}
 \label{tbl:data}
 \begin{tabular}{ccc} 
  \hline
  Obs ID & Date & Net exposure time (ksec)\\
  \hline
  513 & 2000-11-03 &  43 \\
  889 & 2001-09-22 &  50 \\
  5751 & 2006-06-21 & 126 \\
  5752 & 2006-06-21 &  59 \\
  5753 & 2006-06-21 &  74 \\
  6292 & 2006-06-21 &  45 \\
  6295 & 2006-06-21 &  47 \\
  6296 & 2006-06-21 &  49 \\
  \hline
 \end{tabular}
\end{table}

Abell~3667 was observed nine times in total using the {\it Chandra} ACIS-I detectors. We selected eight ObsIDs (513, 889, 5751, 5752, 5753, 6292 and 6295) which have the satellite exposure time of above 10~ksec. We reprocessed the archival standard level 1 event lists produced by the {\it Chandra} pipeline in the standard manner\footnote{CIAO 4.7 Homepage, Data Preparation; http://cxc.harvard.edu/ciao/threads/data.html} using the {\small CIAO} software package (version 4.7) and the {\small CALDB} (version 4.6.5) to apply the appropriate gain maps and the latest calibration products. The net exposure times of each observation after screening are summarized in Table~\ref{tbl:data}. The resulting total net exposure time is $\sim$500~ksec.

We removed bad pixels and also applied the standard data grade selections. We examined the light curve of each observation in the 0.3-10~keV energy band with the standard time binning method recommended in the {\small CIAO} official analysis guides, to exclude periods of anomalously high background. Blank-sky background files provided by the {\it Chandra} team were processed in a similar manner and were scaled by the ratio of the photon counts in the data to those in the background in the high energy band (9.5-12~keV) where the effective area of the telescope is effectively zero.

\subsection{Imaging analysis}
\begin{figure}
  \centering
  \includegraphics[width=3.2in]{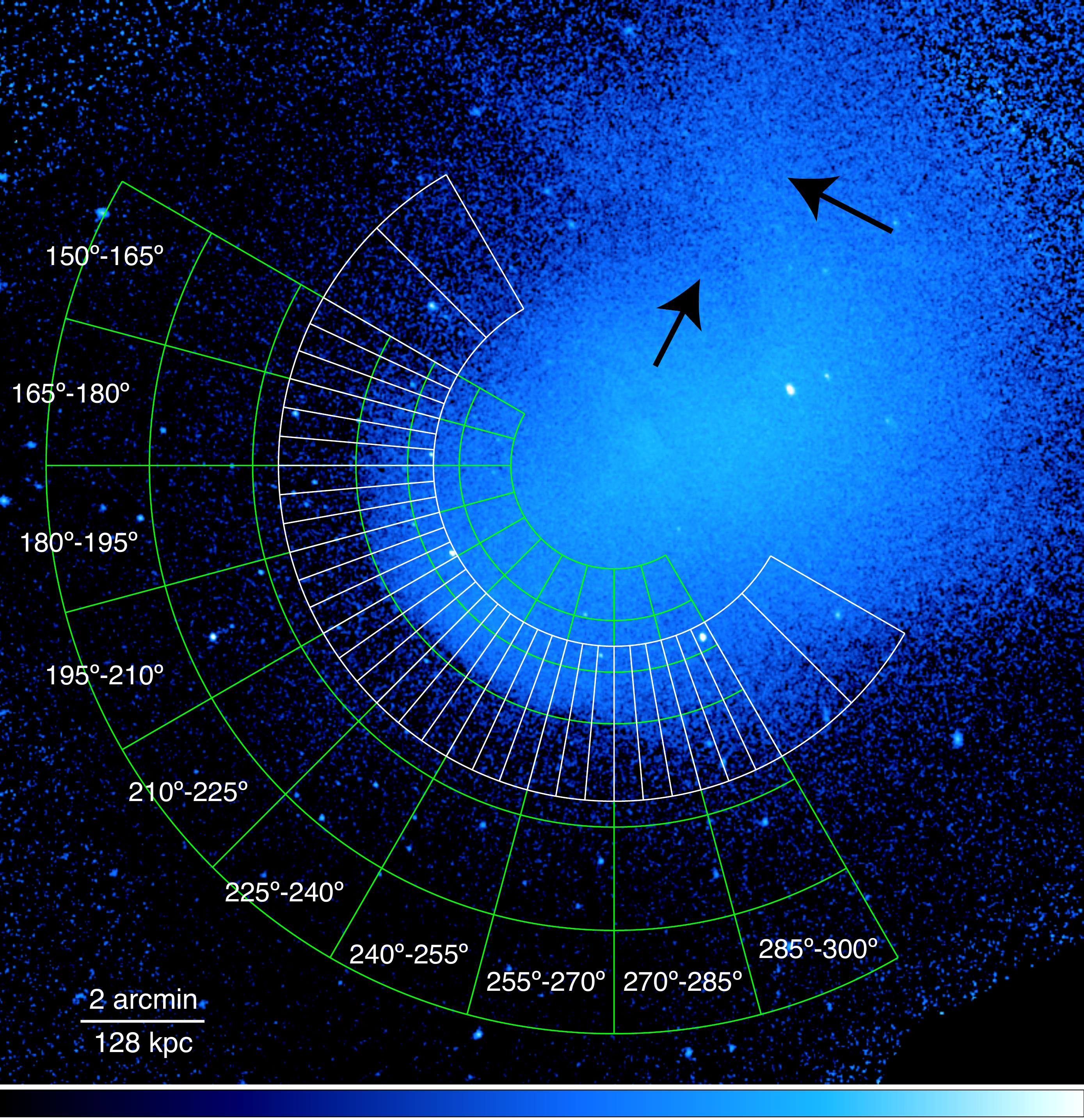}
 \caption[]{$\sigma=0.98$~arcsec Gaussian smoothed, exposure and vignetting corrected, background subtracted {\it Chandra} image (0.6-7.5~keV) of Abell~3667. The overlaid white sectors denote the directions along which the surface brightness profiles are extracted (see also Figures~\ref{img:azimuth_sb} and \ref{img:prof_sb}). The green sectors denote the regions from which the deprojected thermodynamic profiles are extracted (see also Figures~\ref{img:azimuth_thermo}, \ref{img:prof_thermo_1} and \ref{img:prof_thermo_2}). The black arrows indicate the brightness dip and excess (see Section~\ref{sec:global}).}
 \label{img:flatimage}\label{img:img_sb}\label{img:img_deproj}
\end{figure}

We created the exposure and vignetting corrected, background subtracted {\it Chandra} images (flat-fielded image) following the procedure presented by \citet{ichinohe15}. The resulting image is shown in Figure~\ref{img:flatimage}.

\subsection{Surface brightness profiles}
 In order to investigate azimuthal variations of the cold front, we extracted surface brightness profiles for the directions indicated by the white partial-annulus-shaped regions shown in Figure~\ref{img:img_sb}. We determined the centre of these partial annuli so that their radial directions are roughly perpendicular to the cold front. The narrower sectors have an opening angle of 5$^\circ$ and the wider sectors are 15$^\circ$ wide, corresponding to a length of arc of 22~kpc and 67~kpc at the radius of the front.

To model the shape of the surface brightness profiles quantitatively, we assumed that the underlying radial density profile $n(r)$ is expressed as a broken power law with a jump of normalization at the break;
\begin{equation}
 n(r) = \left\{
         \begin{array}{l}
          j_{12}n_0\left(\frac{r}{r_{12}}\right)^{-\alpha_1}\ \ (r \leq r_{12})\\
          n_0\left(\frac{r}{r_{12}}\right)^{-\alpha_2}\ \ (r_{12} < r)
         \end{array}
        \right.,\label{eq:bknpow}
\end{equation}
where $r_{12}$ and $j_{12}$ are the radius of the break and the amplitude of the jump there, $n_0$ is the overall normalization, and $\alpha_1$ and $\alpha_2$ are the power-law slopes of the density profile inside and outside the front.

Ignoring line emission, the emissivity of the ICM is described using the emissivity of thermal bremsstrahlung radiation $\epsilon=\xi(T,Z)n^2$, where $n$ is the density and $\xi(T,Z)$ is a coefficient which weakly depends on temperature and metallicity. Assuming spherical symmetry and approximating $\xi$ to be constant, the surface brightness profile $S(x)$ can be obtained by integrating the density profile along the line-of-sight direction $y$;
\begin{equation}
 S(x) = 2A\int^{\infty}_{0}n^2(r)dy = 4A\int^{\infty}_{0}\frac{x(1+s^2)n(x(1+s^2))}{\sqrt{s^2+2}}ds,
\end{equation}
where $x$ is the coordinate along which the surface brightness profile is extracted, $A$ is a constant which includes both the effect of $\xi$ and the effect of $\propto r^{-2}$ decrement of the intensity and $s$ is a transformed variable using $r=\sqrt{x^2+y^2}=x(1+s^2)$. We fitted this model to the extracted profiles using the {\small Minuit2} fitting library integrated in the {\small ROOT} data analysis framework to minimize $\chi^2$. The results are discussed in Section~\ref{sec:azimuth}.

\subsection{Deprojected thermodynamic profiles}
Thanks to the high-quality deep observation of $\sim$500~ksec, we are able to investigate the deprojected thermodynamic properties with an azimuthal resolution of 15$^\circ$ for a single cold front. Figure~\ref{img:img_deproj} shows in green the 10 directions from each of which we extracted a deprojected thermodynamic profile. For the spectral fitting, we used {\small XSPEC} (version 12.8.2) \citep{arnaud96} to minimize $\chi^2$. The X-ray emission is modeled as a single-temperature thermal plasma using \verb+apec+ model \citep{smith01}. We used the model \verb+projct+ to incorporate the effect of the projection of outer gas volumes, under the assumption of spherical symmetry. We fixed the metallicity to the values obtained from the {\it projected} thermodynamic profiles. We used the chemical abundance table determined by \citet{lodders03}. In the spectral fitting, the typical reduced $\chi^2$ is 1.02, with the typical number of degrees of freedom (NDF) of 1500.

\section{Results}
\subsection{Global morphological features}\label{sec:global}
In the flat-fielded image (Figure~\ref{img:flatimage}), the cold front, the abrupt surface brightness drop that azimuthally extends for $\sim$500~kpc, is clearly visible on the southeastern part of the cluster.

The interface appears abrupt in terms of the drop of the surface brightness. On the other hand, it visually seems to have an azimuthal variation in terms of the radii or the curvature radius, which has not been explicitly pointed out previously. The surface-brightness contrast across the interface is strong towards southeast, but gradually weakened to the north and south along the front, forming a mushroom-like shape.

A brightness dip exists to the north of the cluster's brightness peak, and a brightness excess, which seems to extend anti-clockwise from the brightness peak, exists further out (see the black arrows in Figure~\ref{img:flatimage}). This feature was first mentioned in \citet{mazzotta02} who suggested that this shape is due to the development of hydrodynamic instability.

We also confirmed that the two-dimensional thermodynamic structure, using 500~ksec {\it Chandra} data with contour-binning algorithm \citep{sanders06}, whose bins are themodynamically independent to each other, is consistent with previous observations \citep[][see Appendix~\ref{appendix:thermo}]{mazzotta02,briel04,lovisari09,datta14,hofmann16}.

\subsection{Azimuthally resolved surface brightness properties}\label{sec:azimuth}

\begin{figure*}
  \centering
  \includegraphics[width=6.0in]{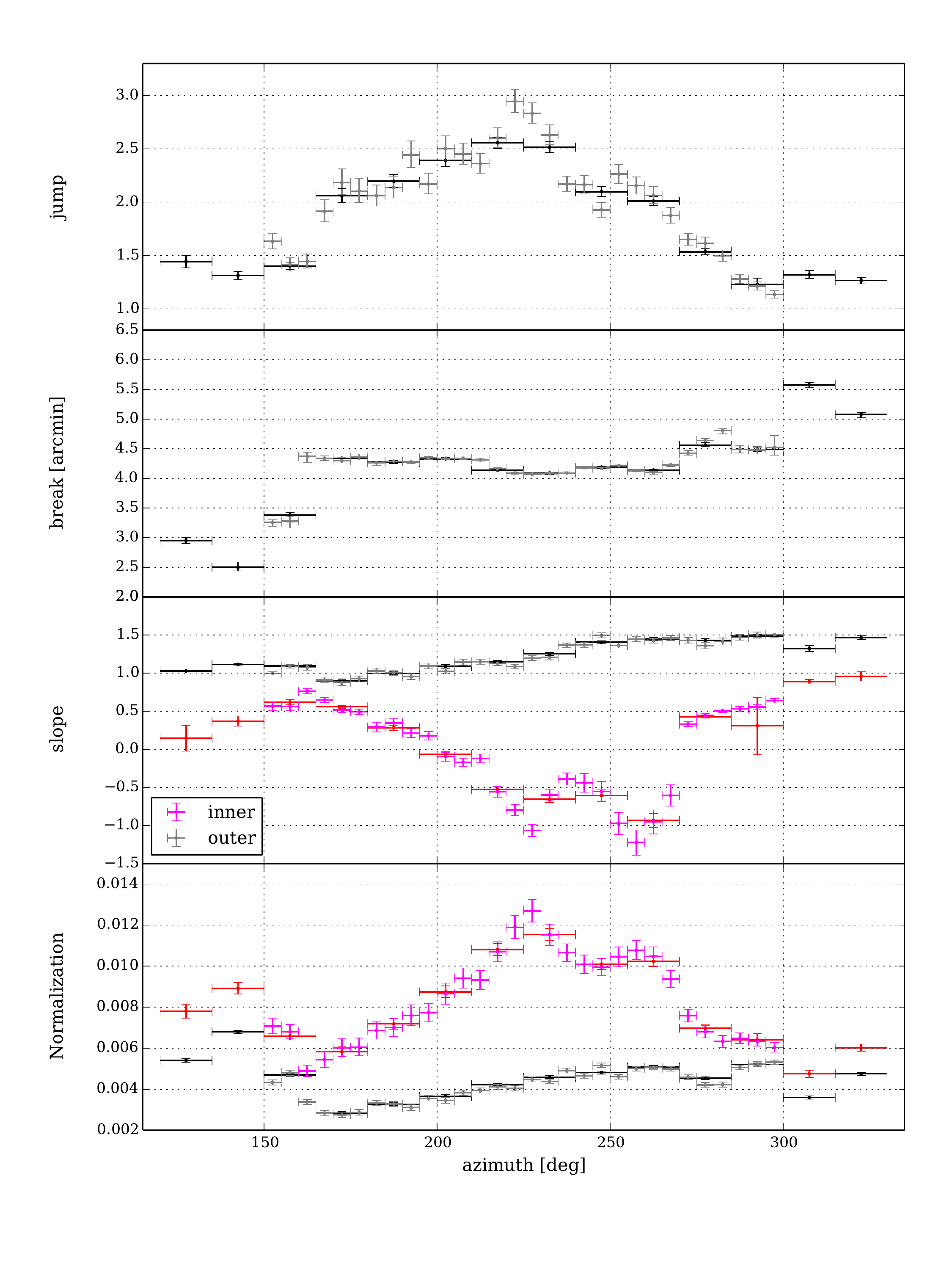}
 \caption[]{The azimuthal variations of the best-fitting parameters of the projected broken power law models. {\it Top row:} the jump $j_{12}$ in density at the break, {\it Second row:} the radius of the break $r_{12}$, {\it Third row:} the slope inside (inner)/outside (outer) the interface ($\alpha_1$, $\alpha_2$). {\it Bottom row:} the normalization inside (inner)/outside (outer) the interface ($j_{12}n_0$, $n_0$). In the bottom two panels, the red/magenta points represent the values inside the break, where the black/gray points represent the values outside the break. The difference between black and gray or between red and magenta is the opening angle of the sectors (15$^\circ$ and 5$^\circ$).}
 \label{img:azimuth_sb}
\end{figure*}
Figure~\ref{img:azimuth_sb} shows the azimuthal variations of the best-fitting parameters of our assumed density model (Equation~\ref{eq:bknpow}). The black/red and the gray/magenta points represent the best-fitting parameters for the surface brightness profile extracted for the wider (15$^\circ$) sectors and the narrower (5$^\circ$) sectors (for the actual surface brightness profiles, in which we can also recognize the azimuthal variations clearly, refer to Figure~\ref{img:prof_sb}).

All of the parameters show significant azimuthal variations. The azimuthal profile of the jump $j_{12}$ is clearly peaked at around 210$^\circ$-240$^\circ$, and has a relatively symmetric shape. The overall trend is the same in the profile of finer azimuthal resolution (5$^\circ$, grey), but there seem to be a systematic bump and a systematic dip of the jump amplitude around 220$^\circ$-225$^\circ$ and 235$^\circ$-250$^\circ$, respectively. The overall trend of the jump, which peaks in the middle of the cold front, is qualitatively consistent with the recent results by \citet{walker16}, who applied the Gaussian gradient magnitude filtering method to the same cold front.

Although the profile of the break $r_{12}$ is more or less constant at azimuths of 170$^\circ$-300$^\circ$, it shows an indication of a downward-convex shape, which may or may not be attributed to the misalignment between the sectors and the cold front. On top of the convex shape, it shows azimuthal variations of relatively large length scale, about 15$^\circ$-25$^\circ$.

The profile of the outer power-law slope $\alpha_1$ shows moderate changes, whereas that of the inner power-law slope $\alpha_2$ shows an asymmetric, two-peaked shape. The inner power-law slopes become positive around the highest contrast part of the front, implying the density increases toward the front.

Although the profile of the outer normalization $n_0$ shows moderate changes, that of the inner normalization $j_{12}n_0$ shows the two-peaked shape similar to that of the inner slope. The difference between the inner normalization and the outer normalization are large at the central azimuths, reflecting the higher contrast of the image there.

\subsubsection{Multiple edges in the surface brightness profiles}
\begin{figure*}
 \begin{minipage}{0.495\hsize}
  \centering
  \includegraphics[width=3.3in]{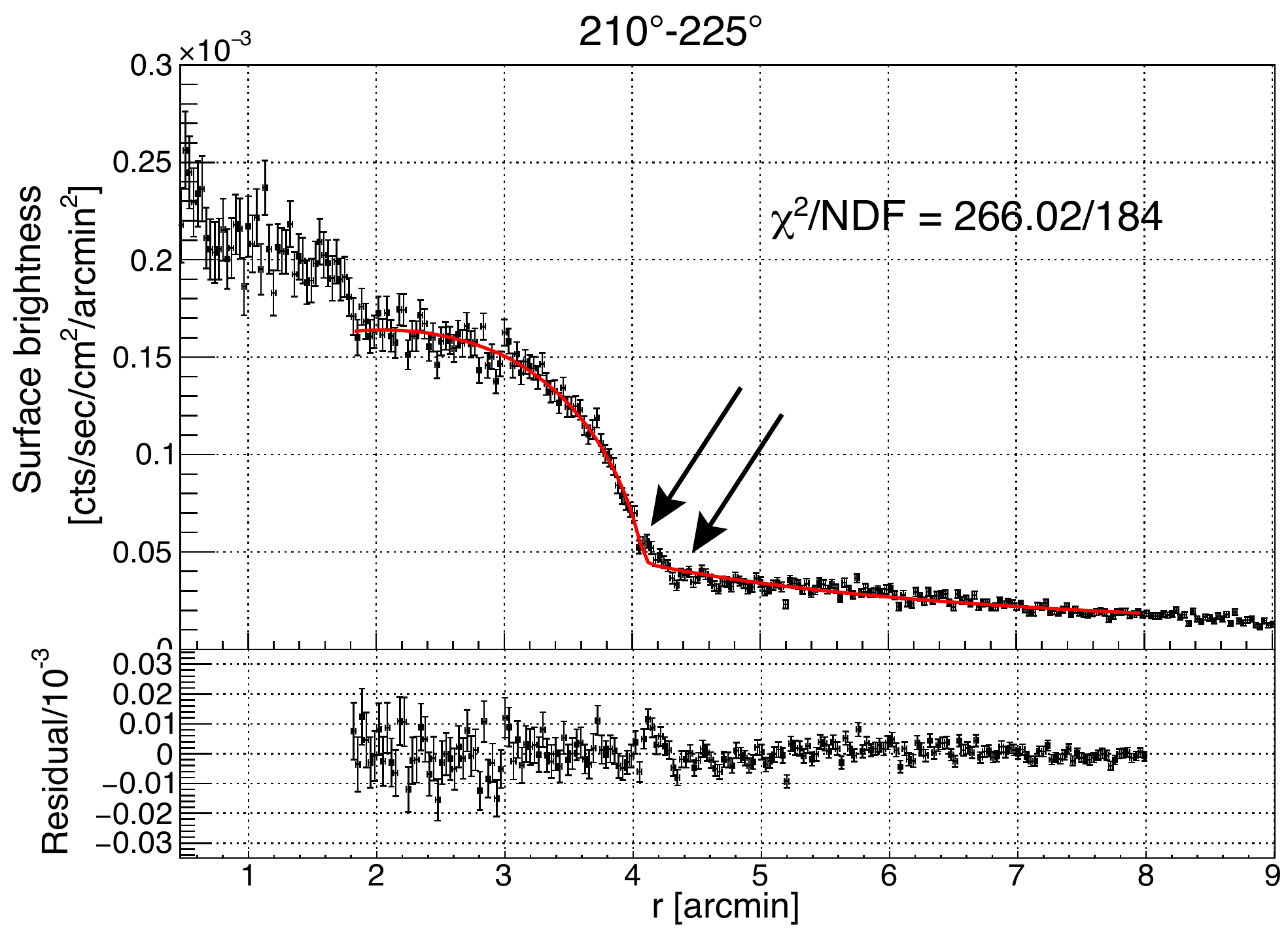}
 \end{minipage}
 \begin{minipage}{0.495\hsize}
  \centering
  \includegraphics[width=3.3in]{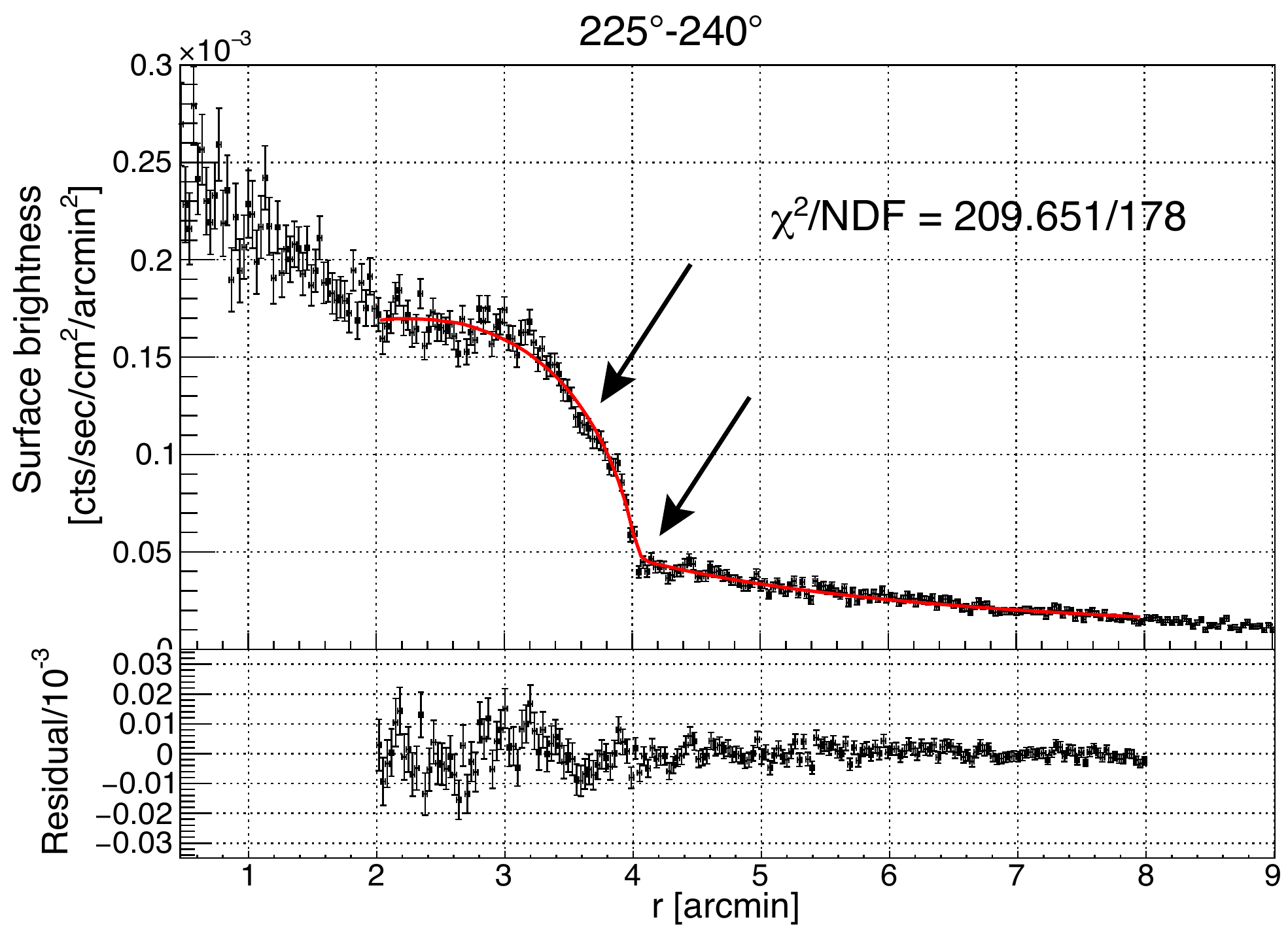}
 \end{minipage}
 \begin{minipage}{0.495\hsize}
  \centering
  \includegraphics[width=3.3in]{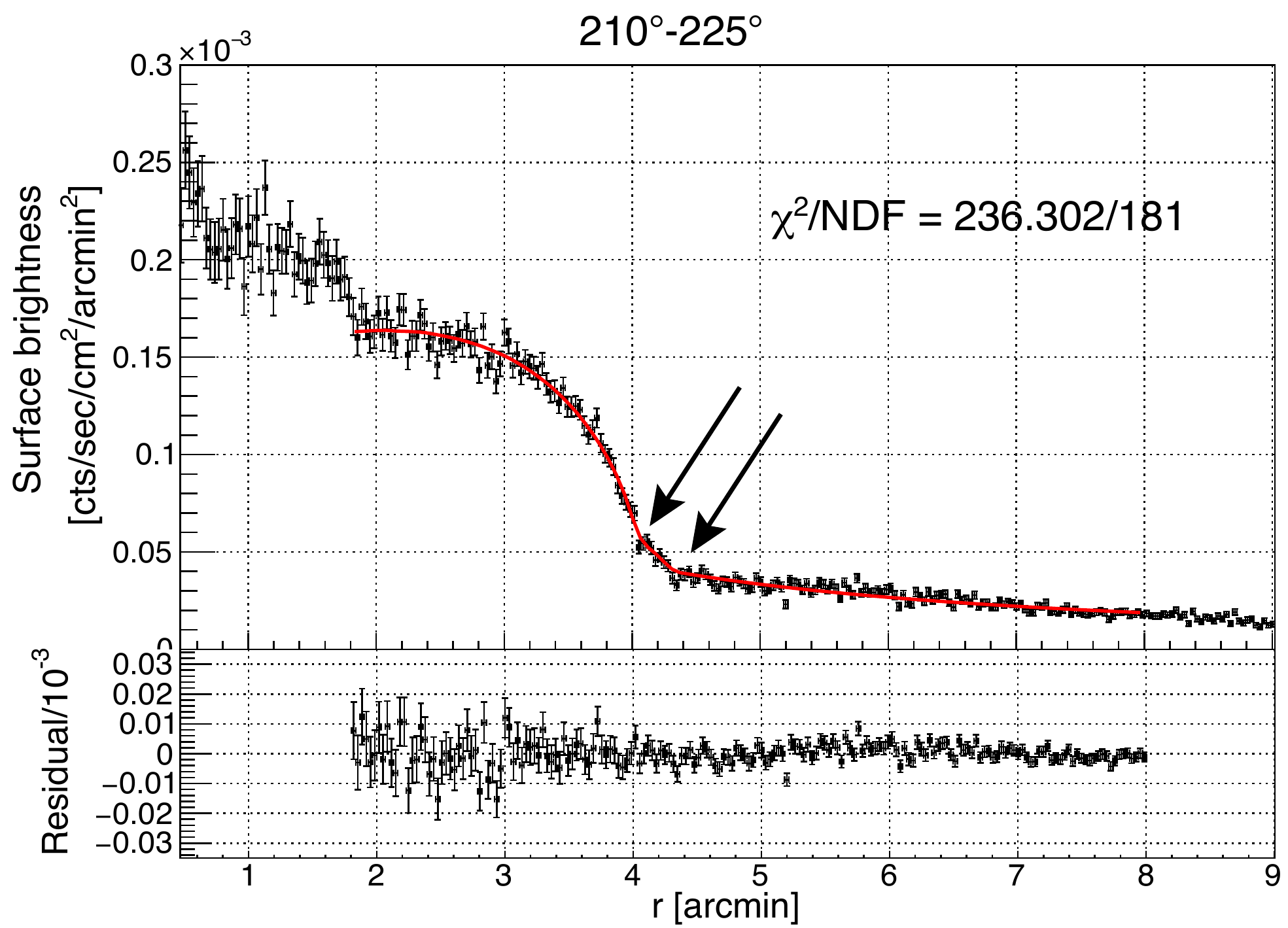}
 \end{minipage}
 \begin{minipage}{0.495\hsize}
  \centering
  \includegraphics[width=3.3in]{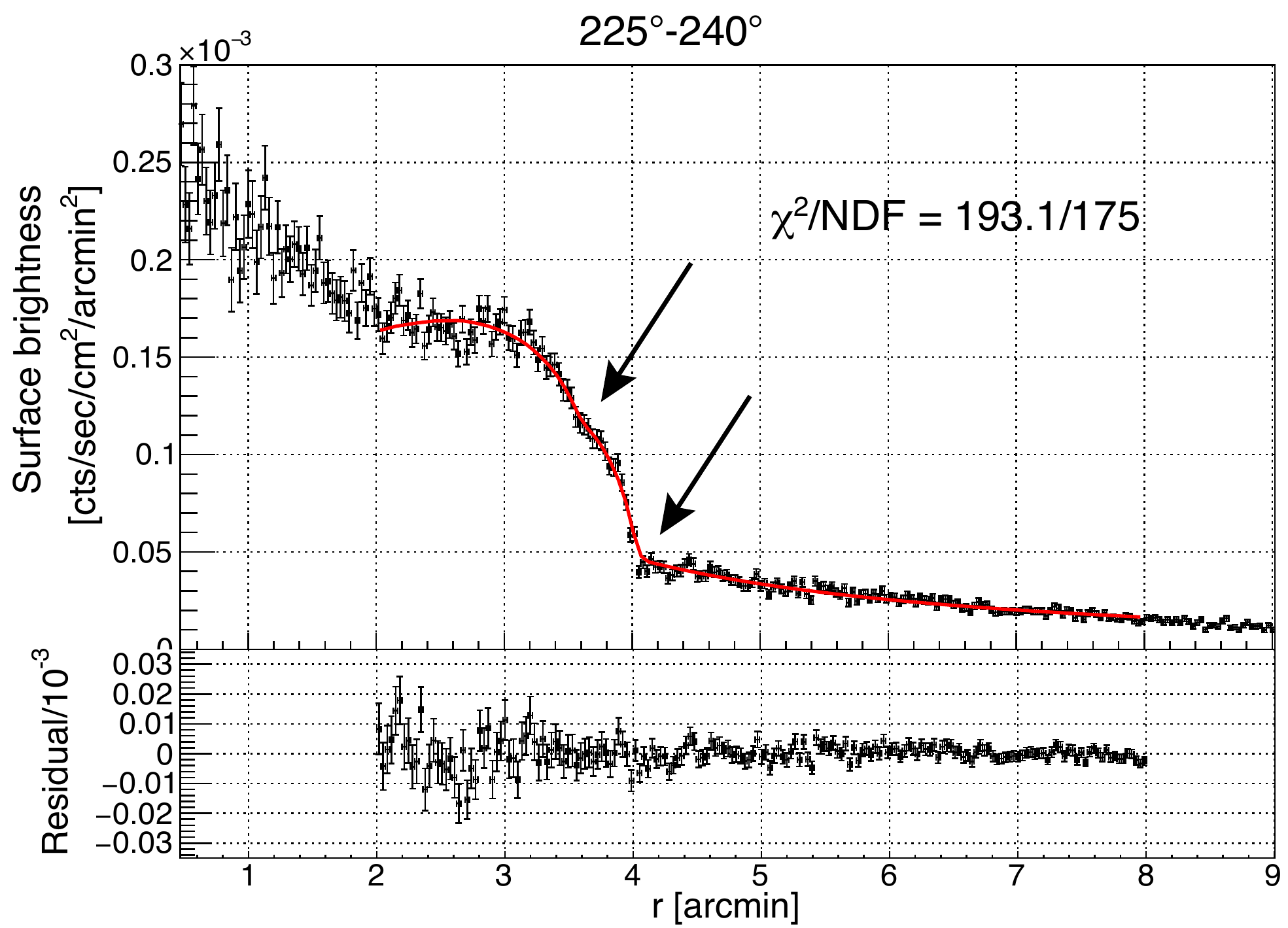}
 \end{minipage}
 \caption[]{The top two panels represent the fitting results for the projected broken power law model (left: 210$^\circ$-225$^\circ$, right: 225$^\circ$-240$^\circ$). In the bottom two panels, the best-fitting projected double-broken power law models are overlaid instead of the best-fitting projected single-broken power law (left: 210$^\circ$-225$^\circ$, right: 225$^\circ$-240$^\circ$). The black arrows represent the positions of the edges.}
 \label{img:prof_sb_dbknpow}
\end{figure*}
Although fitting the surface brightness profiles with a projected broken power law model (Equation~\ref{eq:bknpow}) yields reasonable fits, we see some systematic residuals in the sectors of 210$^\circ$-225$^\circ$ and 225$^\circ$-240$^\circ$ (see the top panels of Figure~\ref{img:prof_sb_dbknpow}). Indeed, each fit improves by $>3\sigma$ level when the surface brightness profiles are fitted with a projected double-broken power law model using the radial density profile of 
\begin{equation}
 n(r) = \left\{
         \begin{array}{l}
          j_{12}j_{23}n_0\left(\frac{r_{12}}{r_{23}}\right)^{-\alpha_2}\left(\frac{r}{r_{12}}\right)^{-\alpha_1}\ \ (r \leq r_{12})\\
          j_{23}n_0\left(\frac{r}{r_{23}}\right)^{-\alpha_2}\ \ (r_{12} < r \leq r_{23})\\
          n_0\left(\frac{r}{r_{23}}\right)^{-\alpha_3}\ \ (r_{23} < r)
         \end{array}
        \right.,\label{eq:dbknpow}
\end{equation}
where $r_{12}$/$r_{23}$ are the radii of the inner/outer breaks, $j_{12}$/$j_{23}$ are the jumps at the inner/outer breaks, $n_0$ is the normalization and $\alpha_1$/$\alpha_2$/$\alpha_3$ are the innermost/middle/outermost power-law slopes ($\Delta\chi^2$ = 29.7 and 16.6 for $\Delta$NDF$=3$ respectively for the sectors 210$^\circ$-225$^\circ$ and 225$^\circ$-240$^\circ$, compared to the single-broken power law model).

The bottom panels of Figure~\ref{img:prof_sb_dbknpow} show the fitting results using the projected double-broken power law model. The residuals are mitigated (see the arrows in Figure~\ref{img:prof_sb_dbknpow}). The density jumps are significant; $j_{12}=1.68_{-0.23}^{+0.31}$ and $j_{23}=1.65_{-0.22}^{-0.23}$ for the 210$^\circ$-225$^\circ$ sector and $j_{12}=1.21\pm0.07$ and $j_{23}=2.59\pm0.10$ for the 225$^\circ$-240$^\circ$ sector.

\subsection{Azimuthally resolved deprojected thermodynamic properties}\label{sec:deproj}
\begin{figure*}
  \centering
  \includegraphics[width=6.0in]{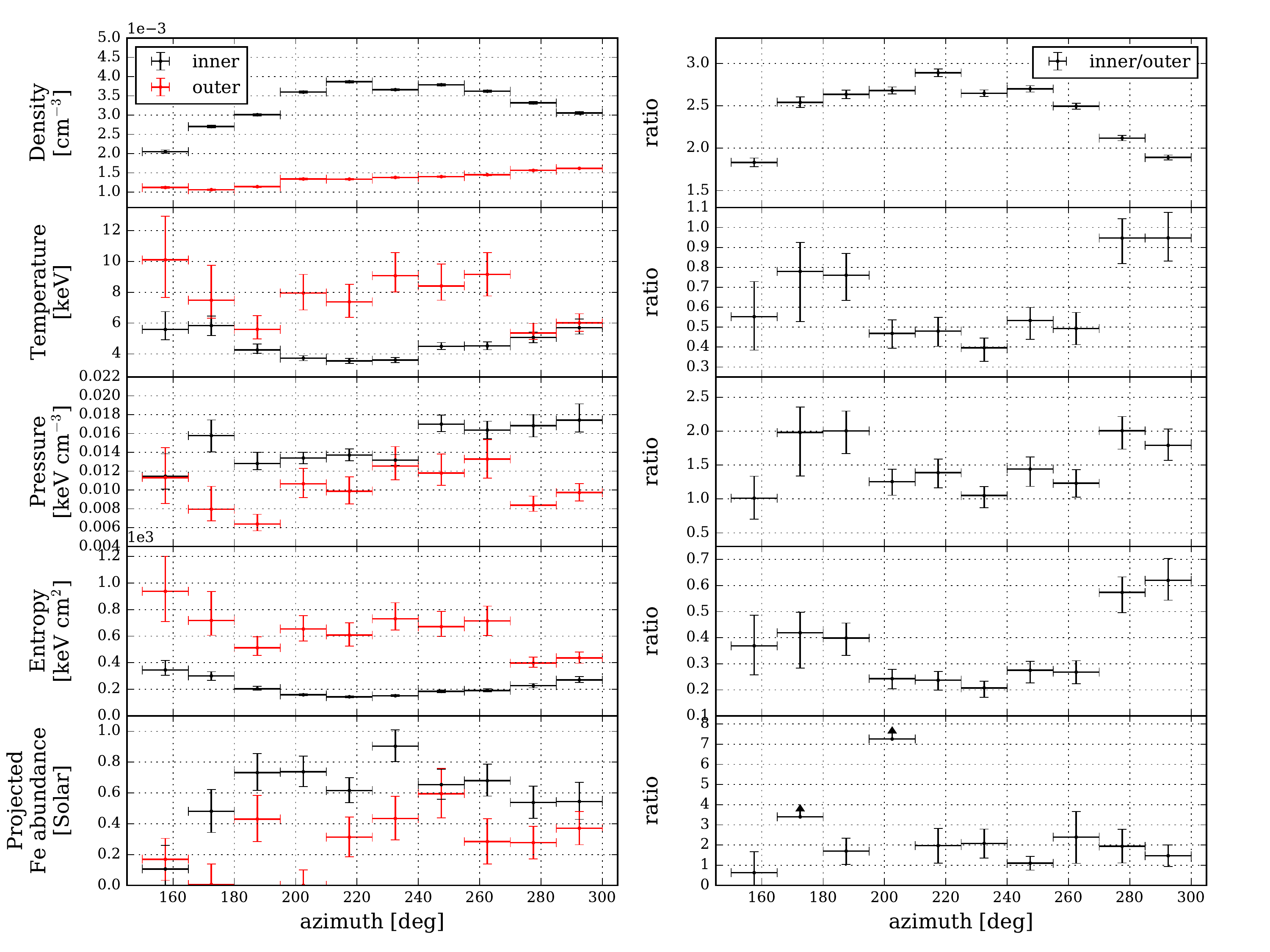}
 \caption[]{The azimuthal variations of the thermodynamic quantities just inside (inner; black) and outside (outer; red) the front. The left panels show the azimuthal variations of the deprojected density, the deprojected temperature, the deprojected pressure, the deprojected entropy and the {\it projected} Fe abundance from top to bottom. The right panels show the ratio of the inner value to the outer value for the corresponding quantities.}
 \label{img:azimuth_thermo}
\end{figure*}

Figure~\ref{img:azimuth_thermo} shows the azimuthal variations of the thermodynamic quantities (the deprojected density, temperature, pressure and entropy, and the {\it projected} Fe abundance) just under and above the interface, together with the ratio of the inner quantity to the outer quantity. For the complete deprojected thermodynamic profiles (from $-150$~arcsec to 300~arcsec), refer to Figures~\ref{img:prof_thermo_1} and \ref{img:prof_thermo_2}.

Generally, at the high-contrast parts of the front (195$^\circ$-270$^\circ$), clearly both the density and the temperature show a jump by a factor of $\sim$2-3. Since the jumps of the density and the temperature are in the opposite sense, the entropy also exhibits a large jump and the pressure shows an almost continuous profile. Underneath the front, the entropy and the temperature profiles seem to show downtrends toward the interface, while the density profiles show uptrends. On the other hand, at the low-contrast part of the front ($<$195$^\circ $ and 270$^\circ<$) the temperatures show milder jumps, resulting in relatively bigger/smaller jumps in the pressure/entropy profiles. These ratios are consistent with the results by \citet{owers09b} who examined the front focusing only on the highest-contrast part (about 210$^\circ$-240$^\circ$). The {\it projected} Fe abundance ratio indicates a slight decrease in the lower-contrast parts, but due to the large errorbars, it is difficult to derive any significant statements.

\section{Discussion}
\subsection{Origin of the cold front}
When a volume of gas, stratified in its gravitational potential well, is subjected to ambient flow, a very sharp cold front will quickly form \citep{markevitch07}. It has been shown in a number of numerical simulations that, in such a situation, the shape of the front will be mushroom-like when it is seen from a direction perpendicular to the flow direction \citep{heinz03,roediger15a,roediger15b}.

In addition to the mushroom-shape in the image, the front also shows a characteristic thermodynamic structure. That is, the dense core gas with low-entropy and high-metallicity is gradually uplifted during the motion, and finally reaches the leading edge of the front. This leads to the thermodynamic structure where the lowest-entropy/temperature and highest-metallicity gas is just below the front.

In our case, as seen in Figure~\ref{img:flatimage}, the shape of the front is clearly mushroom-shaped, and the edges of the front seem to be dissolved into the ambient medium. The deprojected thermodynamic profiles (see Figure~\ref{img:azimuth_thermo}, and also Figures~\ref{img:prof_thermo_1} and \ref{img:prof_thermo_2}) as well as the projected ones in the literature \citep[][see also Figure~\ref{img:thermo}]{mazzotta02,briel04,lovisari09,datta14,hofmann16} clearly display the thermodynamic structure typical of uplifted core gas. In addition, the pressure map which is highly elongated in the northwest-southeast direction, and also a number of observational indications, e.g. the galaxy distribution which is well aligned along the major axis \citep{proust88,owers09a} and the radio relics located to northwest and southeast \citep{rottgering97}, suggest that the cold front is forming because of a merger very close to the sky plane.

\begin{figure}
  \centering
  \includegraphics[width=3.0in]{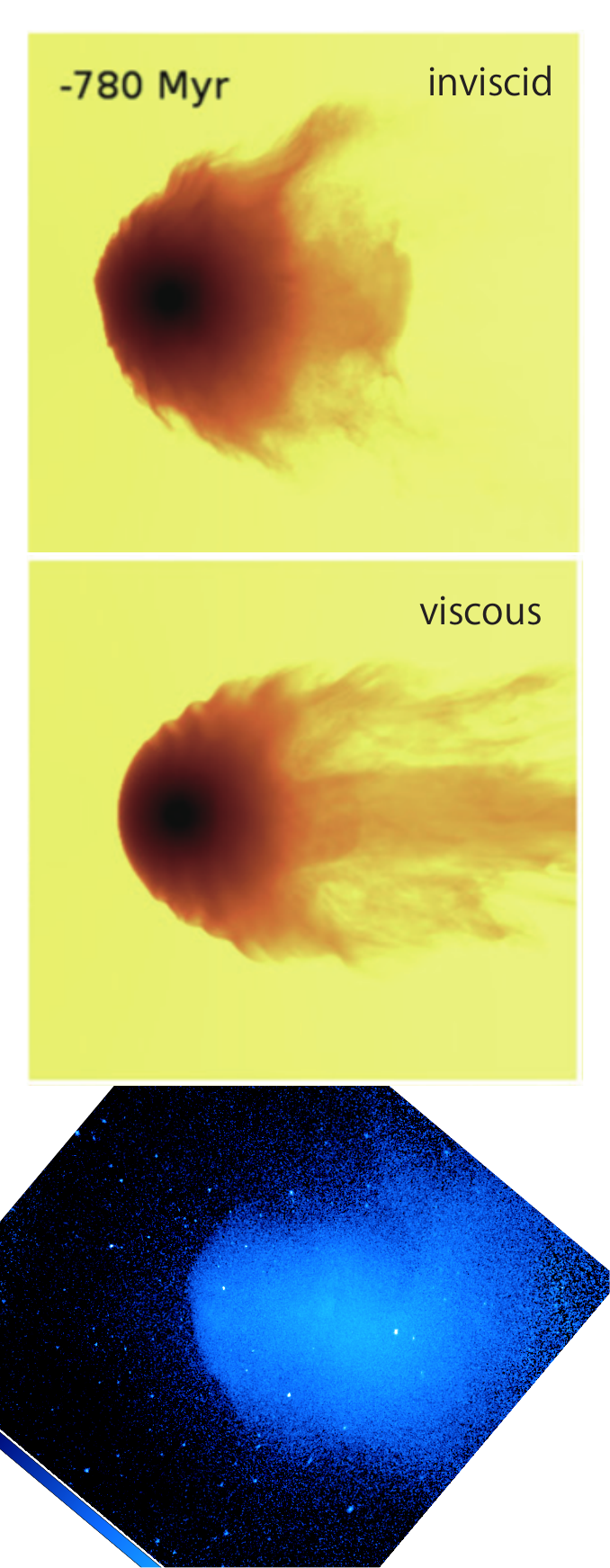}
 \caption[]{{\it Top:} X-ray image (0.7-1.1~keV) of simulated gas-stripped galaxy for the inviscid atmosphere. {\it Middle:} The same image as the top panel but for the viscous (Re$=$46 at pericentre, 0.1~Spitzer viscosity) atmosphere.
 {\it Bottom:} Figure~\ref{img:flatimage}, $40^\circ$ clockwise rotated. The size of each simulation box is $\sim$100~kpc, while that of the X-ray image is $\sim$1.1~Mpc.
The top and the middle panels are reproduced from Figure~6 (third column, fourth row panel) and Figure~8 (third column, fourth row panel) in \citet{roediger15b}, respectively\protect\footnotemark.
 }
 \label{img:roediger15}
\end{figure}
\footnotetext{Roediger E., et al., ``Stripped Elliptical Galaxies as Probes of ICM Physics: II. Stirred, but Mixed? Viscous and Inviscid Gas Stripping of the Virgo Elliptical M89'', The Astrophysical Journal, 806:104, 15pp. (2015 June 10). \copyright AAS. Reproduced with permission.}

 We also point out the striking similarity of the X-ray image to the numerical simulation by \citet{roediger15b} who modeled the inviscid stripping of an initially extended atmosphere subjected to the ambient flow of its host cluster during the initial relaxation phase (Figure~\ref{img:roediger15}). The cold front in Abell~3667 (Figure~\ref{img:roediger15} bottom) is especially similar to the inviscid simulation result at 780~Myr before the pericentre passage (Figure~\ref{img:roediger15} top) in two perspectives; (1) the opening of the front or the angle where the stripping starts, and more importantly, (2) the variations of the front radii whose length scale is smaller than the opening angle of the entire cold front (sub-opening-angle scale variation).

It is also shown in the numerical simulations by \citet{roediger15b} that the impact of the inclination angle of the line-of-sight direction with respect to the direction perpendicular to the merger plane is relatively strong above $\sim$30$^\circ$, for which case the interface becomes less pronounced and the sub-opening-angle scale variations are no longer clearly visible. 

 Given all the arguments above, we suggest that the front is formed via a merger event taking place nearly in the sky plane. Note that gas sloshing parallel to the line of sight direction is another possible interpretation \citep{kitayama14}. Testing these different scenarios would require measurements of the line-of-sight velocity inside and outside the interface by Doppler shift measurements, which can only be performed with high resolution X-ray spectroscopy.

\subsection{Gas dynamics}
From the thermodynamic information, we can estimate the velocity of the cool gas relative to the ambient medium as has been done in the literature \citep{vikhlinin01b, landau}: by approximating the cool gas as a blunt body subjected to an ambient flow, and neglecting the change in the gravitational potential along the streamline, the ratio of the pressure of the flow at the stagnation point $p_0$ to the pressure of the flow in the free streaming region $p_1$ is a function of the cloud velocity $v$;
\begin{equation}
 \frac{p_0}{p_1}
  = \left\{
     \begin{array}{l}
      \left(1+\frac{\gamma-1}{2}{M_1}^2\right)^{\gamma/(\gamma-1)}\ \ (M_1 \leq 1)\\
      \left(\frac{\gamma+1}{2}\right)^{(\gamma+1)/(\gamma-1)}{M_1}^2\left(\gamma-\frac{\gamma-1}{2{M_1}^2}\right)^{-1/(\gamma-1)}\ \ (M_1 > 1)
     \end{array}
    \right.,
\end{equation}
 where $\gamma = 5/3$ is the adiabatic index of the monoatomic gas and $M_1 = v/c_s$ is the Mach number of the free stream, with $c_s$ being the free-stream sound speed.

Assuming that the azimuthal range of 210$^\circ$-240$^\circ$ represents the tip of the front and using $p_0 \sim 0.0134\pm0.0004~\mr{keV~cm^{-3}}$ (average of the bins just below the front) and $p_1\sim 0.0092_{-0.0004}^{+0.0005}~\mr{keV~cm^{-3}}$ (average of the outermost bins) from the deprojected thermodynamic profiles (see also Figures~\ref{img:prof_thermo_1} and \ref{img:prof_thermo_2}), the Mach number of the free stream is calculated at $M_1 = 0.70\pm0.06$. This value is consistent with the previous estimation by \citet{datta14} but smaller than the previous estimation of 1.0$\pm$0.2 by \citet{vikhlinin01b}; this discrepancy may be due to the indirect measurement of $n_e$ by \citet{vikhlinin01b}, who inferred the density assuming that the surface brightness profile follows a $\beta$ model.

The sound speed $c_{s1}$ in the free stream is calculated using $c_{s1} = \sqrt{\gamma kT_1/\mu m_p}$, where $kT_1$ is the temperature of the free stream, and $\mu=0.6$ is the mean particle weight with respect to the proton mass $m_p$. Using $kT_1 = 6.9_{-0.3}^{+0.4}~\mr{keV}$ and $M_1 = v/c_{s1}$, the velocity of the cool gas is estimated at $v = 950\pm80~\mr{km~s^{-1}}$.

\subsection{Kelvin-Helmholtz instability}
\subsubsection{Sub-opening-angle scale variations}
What makes the cold front in Abell~3667 a particularly interesting target are its variations on azimuthal scales smaller than the opening angle of the entire front (sub-opening-angle scale variation). This variation may have been missed in other cold fronts because the studies done so far have focused mainly on the width of the front and the surface brightness profiles have been extracted from much wider sectors \citep[e.g.][]{owers09b,ghizzardi10,datta14}.

As we pointed out in the previous Section, the radii of the front show azimuthal variations, and the variation is similar to the inviscid simulation results (Figure~\ref{img:roediger15} top panel). Actually, high-viscosity (Re$=$46 at pericentre, 0.1~Spitzer viscosity) simulation results are qualitatively inconsistent with our observations because they predict much smoother interfaces (Figure~\ref{img:roediger15} middle panel). This suggests that the Reynolds number of the ICM is much higher in Abell~3667.

\citet{roediger15b} suggest that, in the inviscid stripping, the momentum transfer between two gas phases occurs via Kelvin-Helmholtz instabilities (KHIs), while in the viscous case, it does via viscosity. The simulation results (Figure~\ref{img:roediger15} top panel) clearly show KHIs occurring close to the tip of the front and developing toward the edge, strongly suggesting that our azimuthal variations of the interface are the signatures of the onset of developing KHIs. We investigate the scenario more in detail later in Sections~\ref{sec:comp} and \ref{sec:viscosity}.

\subsubsection{Multiple edges}\label{sec:multipleedge}
Recently it has been suggested by \citet{roediger13a} that, when there are developing KHIs at a cold front, the surface brightness profile across the front exhibits multiple edges, similarly to our case. The differences in radii between the breaks are $\Delta r_\mr{break} \equiv r_{23}-r_{12} = 0.26_{-0.04}^{+0.02}$~arcmin and $\Delta r_\mr{break} = 0.46_{-0.05}^{+0.03}$~arcmin respectively for the sectors 210$^\circ$-225$^\circ$ and 225$^\circ$-240$^\circ$, which correspond to the actual lengths of $\Delta r_\mr{break} = 17_{-3}^{+1}~\mr{kpc}$ and $\Delta r_\mr{break} = 30_{-3}^{+2}~\mr{kpc}$.

\citet{roediger13a,roediger13b} further suggested that the separation between the edges corresponds to about a fourth to a half of the scale length of the KH rolls. From $\Delta r_\mr{break}$ values, the scale length of the KH rolls is thus estimated at around 30-120~kpc, which is consistent with the observed sub-opening-angle scale variation of the interface (see also Section~\ref{sec:comp}).

\begin{figure*}
 \begin{minipage}{0.495\hsize}
  \centering
  \includegraphics[width=3.0in]{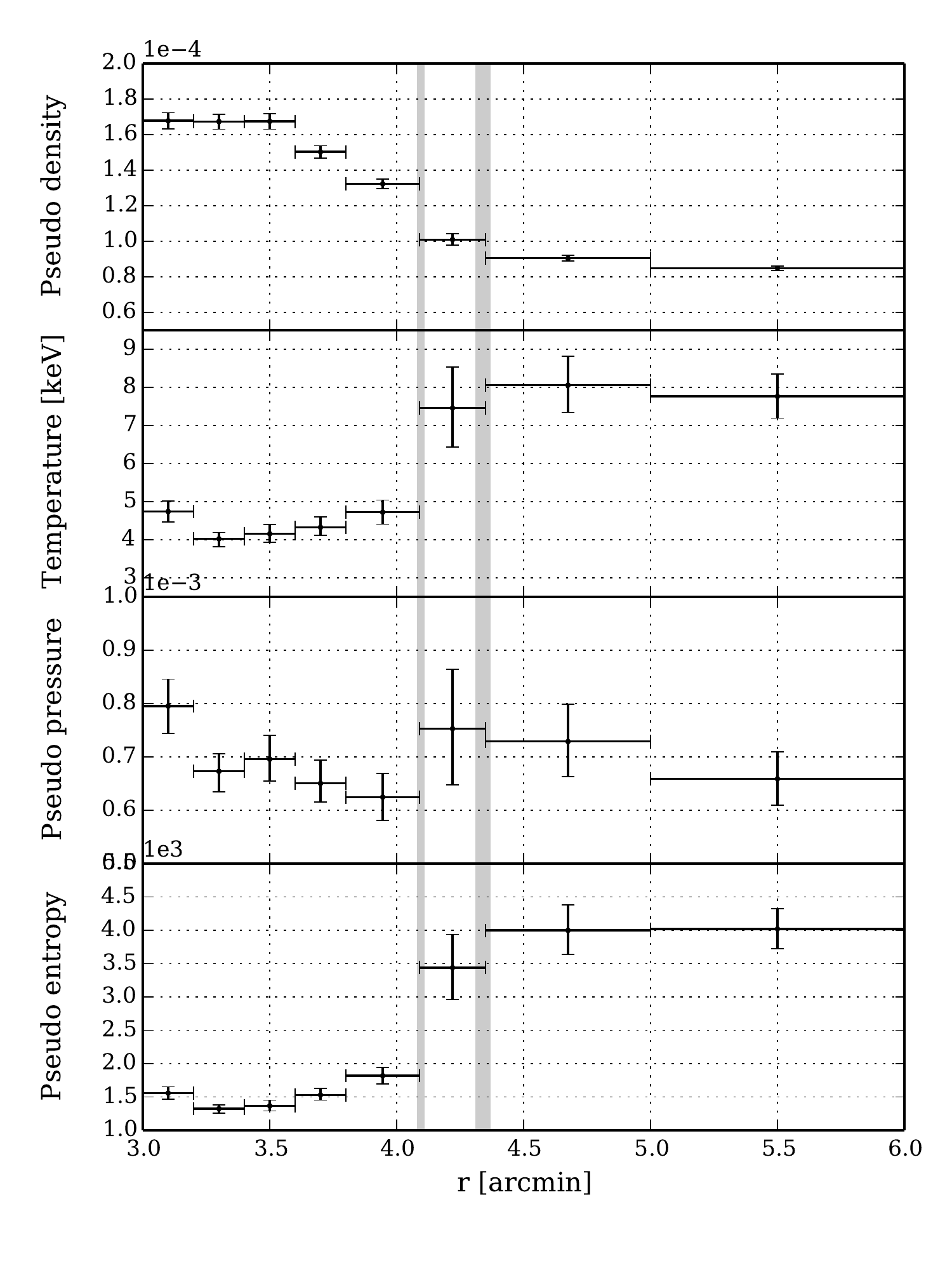}
 \end{minipage}
 \begin{minipage}{0.495\hsize}
  \centering
  \includegraphics[width=3.0in]{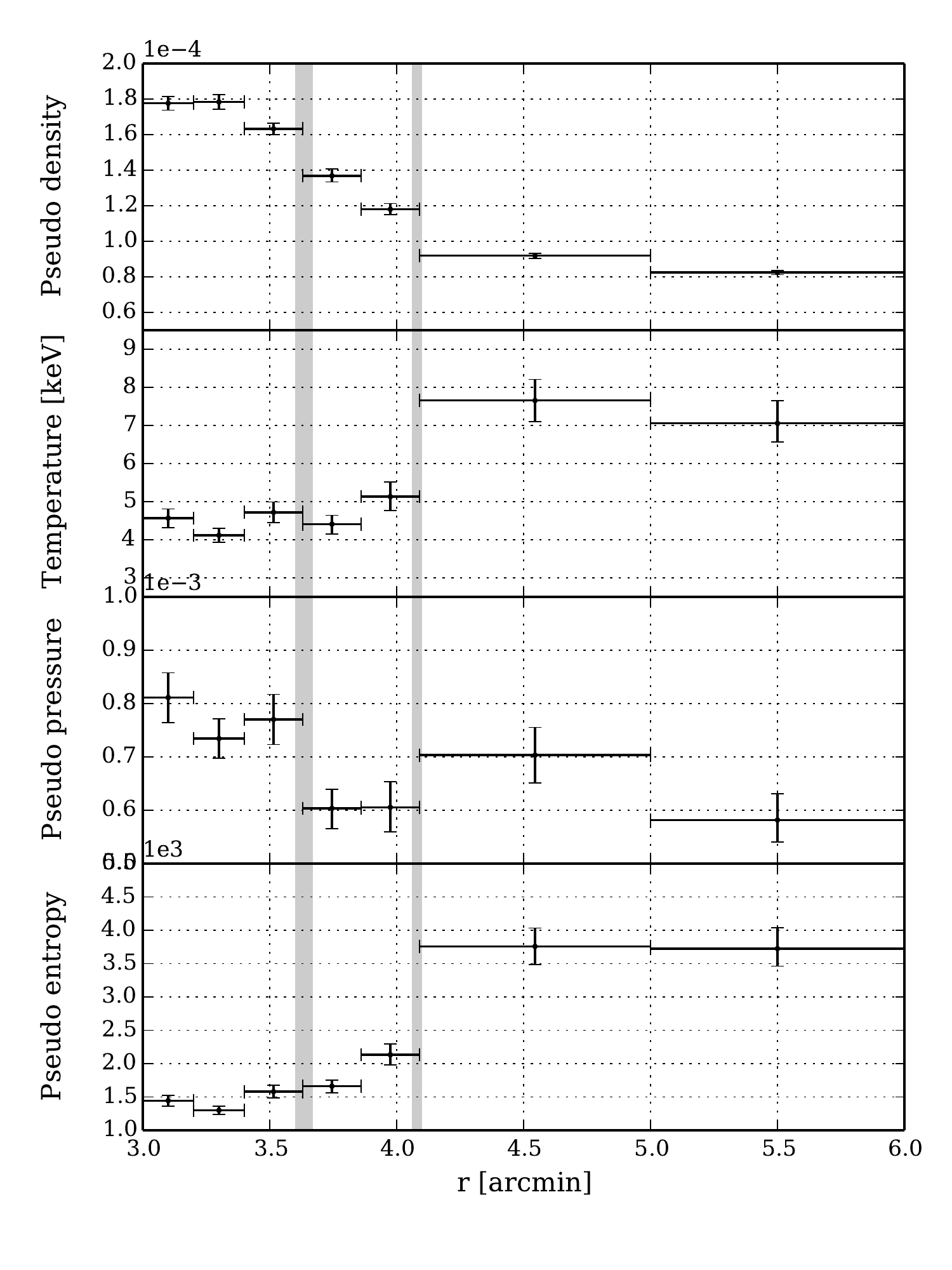}
 \end{minipage}
\cprotect\caption[]{{\it Projected} thermodynamic profiles in the two directions where we see multiple edges in the corresponding surface brightness profiles ({\it Left:} 210$^\circ$-225$^\circ$ and {\it Right:} 225$^\circ$-240$^\circ$). The panels are the pseudo density (calculated using $\tilde{n}=\sqrt{N/A}$ where $N$ is the \verb+apec+ normalization and $A$ is the area of the spectral extraction region in pixel$^2$), the temperature ($kT$), the pseudo pressure ($\tilde{n}kT$) and the pseudo entropy ($kT\tilde{n}^{-2/3}$) from top to bottom. The vertical gray bands denote the edges in the surface brightness profiles.}
 \label{img:prof_multiple}
\end{figure*}

Figure~\ref{img:prof_multiple} shows the {\it projected} thermodynamic profiles for the corresponding sectors. In the profiles extracted for the sector 210$^\circ$-225$^\circ$ (Figure~\ref{img:prof_multiple} left), we see a jump of the temperature and entropy at the inner edge while they are continuous at the outer edge, which means that the thermodynamic properties between the edges are similar to those of the outer ambient gas. Given that the break radii are different between the sectors 210$^\circ$-215$^\circ$ and 215$^\circ$-225$^\circ$ as shown in Figure~\ref{img:azimuth_sb}, it is likely that the multiple edges in this direction are caused by the fluctuation of the break radii and the thermodynamic properties between the edges are dominated by the outer ambient gas.

In contrast, in the profiles extracted for the sector 225$^\circ$-240$^\circ$ (Figure~\ref{img:prof_multiple} right), we see a jump of the temperature and entropy at the outer edge while they are continuous at the inner edge. Moreover, the pressure profile seems to exhibit a deficit between the edges. As shown in Figure~\ref{img:azimuth_sb}, the break radii measured in the 15$^\circ$ resolution are consistent with the break radii measured in the 5$^\circ$ resolution. This means that both the multiple edges and the thermodynamic structure in this direction are not attributable to the simple azimuthal resolution effect unlike the case of the 210$^\circ$-225$^\circ$ sector (previous paragraph), strongly indicating the existence of a projected KHI layer.

Under the KHI scenario, the pressure deficit can be explained: as KHI eddies collapse into smaller eddies and ultimately into turbulence, the turbulent pressure may support the total pressure, lowering the apparent gas pressure. The deficit of the pseudo pressure of $\Delta\tilde{p}\sim0.1\times10^{-3}$ is translated to the physical pressure deficit of $\Delta p\sim4.4\times10^{-3}~\mr{keV cm^{-3}}\times(l/50~\mr{kpc})^{-1/2}$, where $l$ is the line-of-sight depth of the structure. Assuming the pressure deficit is caused by the KHI turbulence, we estimate the typical turbulent strength at $v\sim350~\mr{km/s}$, using $\Delta p = \rho v^2$, where $\rho$ is the mass density and $v$ is the turbulent velocity dispersion. This value is smaller than the typical shear strength at the interface and is compatible with the above mentioned scenario.

All these measurements are consistent with the scenario that KHIs are developing on the interface. Recently the existence of KHIs has been suggested on the sloshing cold fronts in the Centaurus and the Ophiuchus clusters \citep{sanders16,werner16}, and also on the merger cold front in the NGC~1404 galaxy \citep{su16}.  That said, to our knowledge, this is the first observational indication of KHIs near the tip of a merger cold front in a galaxy cluster.

\subsubsection{Gas mixing at the mushroom edge}
As shown in Figure~\ref{img:azimuth_thermo}, the temperature and the entropy jumps are relatively moderate at the edge of the front (i.e. 165$^\circ$-195$^\circ$, 270$^\circ$-300$^\circ$) compared with the tip of the front (i.e. 195$^\circ$-270$^\circ$). Given that these edge azimuths correspond to the edge of the mushroom-shape in the image, this difference is probably due to the mixing of the gas induced by the fully developed, turbulent KHIs. In the case of KHI-induced gas mixing, the Fe abundance is also expected to show a moderate change at the edge compared to the tip. However, we don't see any significant difference along the interface in the {\it projected} Fe abundance profile shown in the bottom panel in Figure~\ref{img:azimuth_thermo}. Note that in the {\it projected} Fe abundance map shown in Figure~\ref{img:thermo} (which has higher statistics per each bin), we see that the contrast of the Fe abundances across the interface seems to be stronger near the tip of the front than at the edge of the front.

\subsubsection{Detailed view of the fluctuation of break radii}\label{sec:comp}
The length scales of the fluctuation of the break radii appear to be limited in the range of 10$^\circ$-25$^\circ$ when we look at the azimuthal profile in the 5$^\circ$ resolution (see gray points in Figure~\ref{img:azimuth_sb}). Here we investigate the properties of the fluctuation by examining the azimuthal profile with finer angular resolutions.

\begin{figure*}
  \centering
  \includegraphics[width=6.0in]{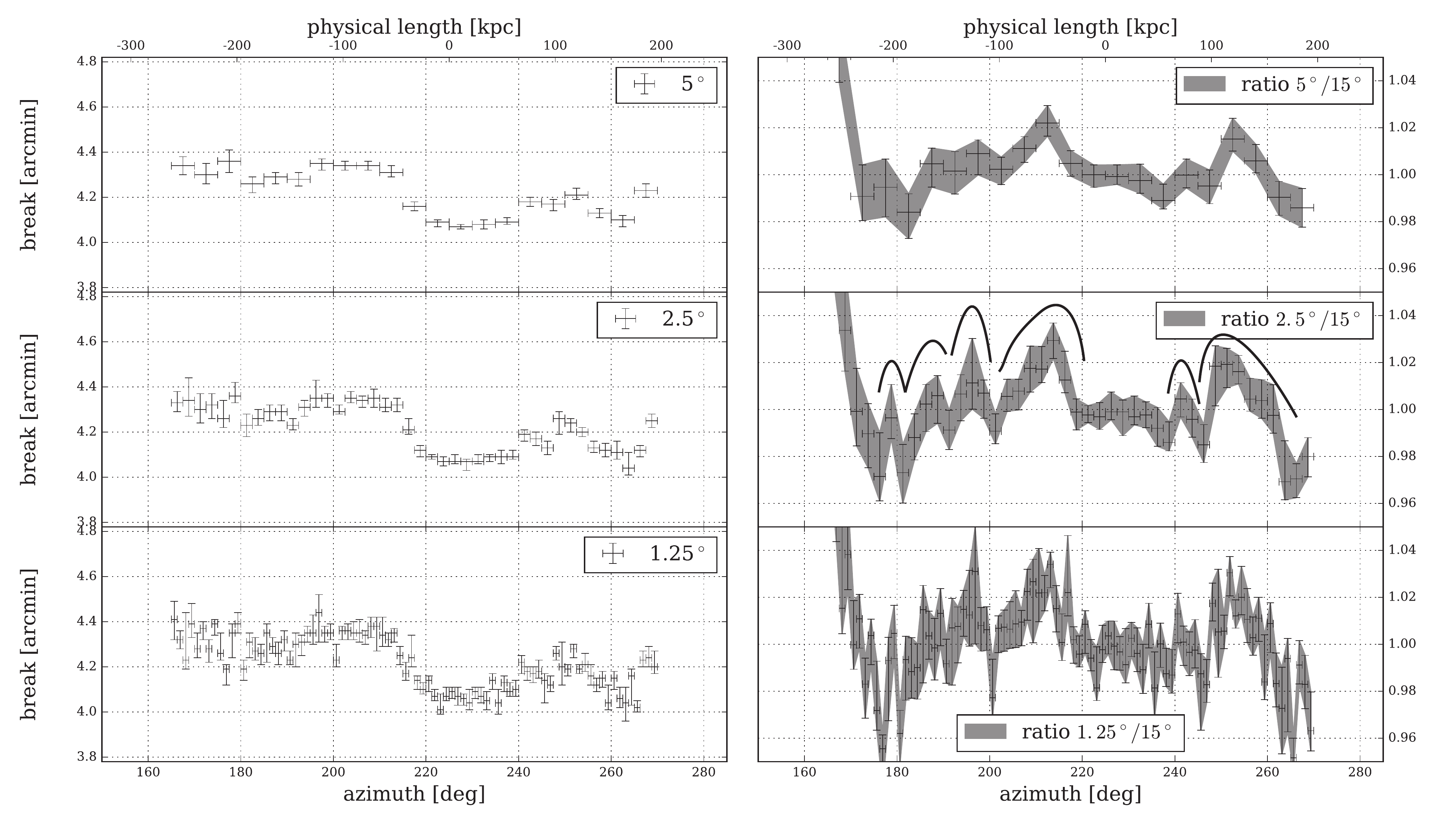}
 \caption[]{{\it Left:} the break radii shown in different azimuthal resolution. 5$^\circ$, 2.5$^\circ$ and 1.25$^\circ$ from top to bottom. {\it Right:} relative fluctuation created by dividing each corresponding left panel by the spline-interpolated break radii profile in 15$^\circ$ resolution. The black curves denote the ``sequential bumps'' (see the text below). The $x$-axes on the top are the physical length along the cold front with the origins corresponding to 225$^\circ$.
 }
 \label{img:comp}
\end{figure*}
Figure~\ref{img:comp} left shows the azimuthal profile of the break radii (165$^\circ$-270$^\circ$) extracted in 5$^\circ$, 2.5$^\circ$, and 1.25$^\circ$ resolutions (see also the second panel of Figure~\ref{img:azimuth_sb} for the 5$^\circ$ and 15$^\circ$ profiles). We find that the overall shapes of the profiles are similar to each other. However, in the finer resolution profiles, we find substructures which are missed even in the 5$^\circ$ profile.

The break radii profiles may be affected by e.g. the misalignment between the normal of the interface and the direction of the sectors from which the profiles are extracted. In order to mitigate such effects and focus on the fluctuations of the break radii themselves, we computed the relative fluctuation profiles by dividing each profile by the spline-interpolated profile using an opening angle of 15$^\circ$. Figure~\ref{img:comp} right shows the azimuthal profiles of the relative fluctuation of the break radii for the corresponding left panels.

In the 2.5$^\circ$ resolution profile, we see sequential bump-like structures, which are missed in the 5$^\circ$ profile (see the curves in Figure~\ref{img:comp}). In the 1.25$^\circ$ profile, we also see the similar sequential bump-like structures, although they are less prominent due to the larger errorbars. The fact that the bumps are less prominent in this resolution indicates that the data quality is insufficient for $\lesssim1.25^\circ$ resolutions. The characteristic size of the sequential bumps is 20-80~kpc, which is similar to the value inferred from the differences of the radii of the breaks in the surface brightness profiles (30-120~kpc; see Section~\ref{sec:multipleedge}).

\subsubsection{Upper limit of the ICM effective viscosity}\label{sec:viscosity}
Assuming that KHIs are actually developing on the interface and that the scales of the fluctuations (sequential bumps) represent the length scales of the KHI eddies, we can extract implications for the effective viscosity of the ICM.

In general, finite shear flow induces exponentially developing KHIs. If the gas were inviscid and incompressible, KHIs would develop on all length scales. However, the growth of the perturbation is suppressed when the Reynolds number Re of the ICM in the hot layer (outside the interface) is below the critical value
\begin{equation}
 \mr{Re} = \frac{\rho\lambda V}{\mu} < \mr{Re}_\mr{crit} \sim 64 \sqrt{\Delta},\label{eq:visc}
\end{equation}
where $\rho$ is the density, $\lambda$ is the length scale, $V$ is the shear strength, $\mu$ is the viscosity, $\rho_1$ and $\rho_2$ are the densities of the two gas phases on the two sides of the interface, and $\Delta = (\rho_1+\rho_2)^2/\rho_1\rho_2$ \citep{roediger13b,chandrasekhar61}. Note that \citet{roediger13b} estimated the value of 64 with a simple analytic argument. They offered some more conservative estimations of $\mr{Re}_\mr{crit}=10\sqrt{\Delta}$ or 16$\sqrt{\Delta}$, but given that we do not see developing KHIs in Figure~\ref{img:roediger15} middle ($\mr{Re}=46$), we think these critical Reynolds numbers are too conservative\footnote{The Reynolds number of 46 is the pericentre value and not exactly the value on the spot at $-780$~Myr. However, the density, size, and velocity on the spot at $-780$~Myr are respectively $\sim$1/3 times, $\gtrsim$6 times, and $\gtrsim$1/2 times the values at the pericentre, and we think it is safe to use 46 as the Reynolds number on the spot in the above argument.}.

This relation means that for a fixed viscosity, the difference of the shear strengths results in the difference of the length scales of the KHI modes which can develop. In other words, when a value of the effective viscosity $\mu$ is given, the perturbation of length-scale $\lambda$ is suppressed if $\lambda < \lambda_\mr{crit} = \mu\mr{Re}_\mr{crit}/\rho V$. 

When a sphere is subjected to incompressible ideal flow, the speed of the fluid on the sphere $v(\theta)$ is expressed using $v(\theta) = 3V\sin\theta/2$, where $V$ is the speed of the flow and $\theta$ is the angle between the flow and the direction of the radius vector of a given position on the sphere. Therefore, assuming that the interface is spherical, we can calculate the flow speed at each azimuth.

\begin{figure*}
  \centering
  \includegraphics[width=6in]{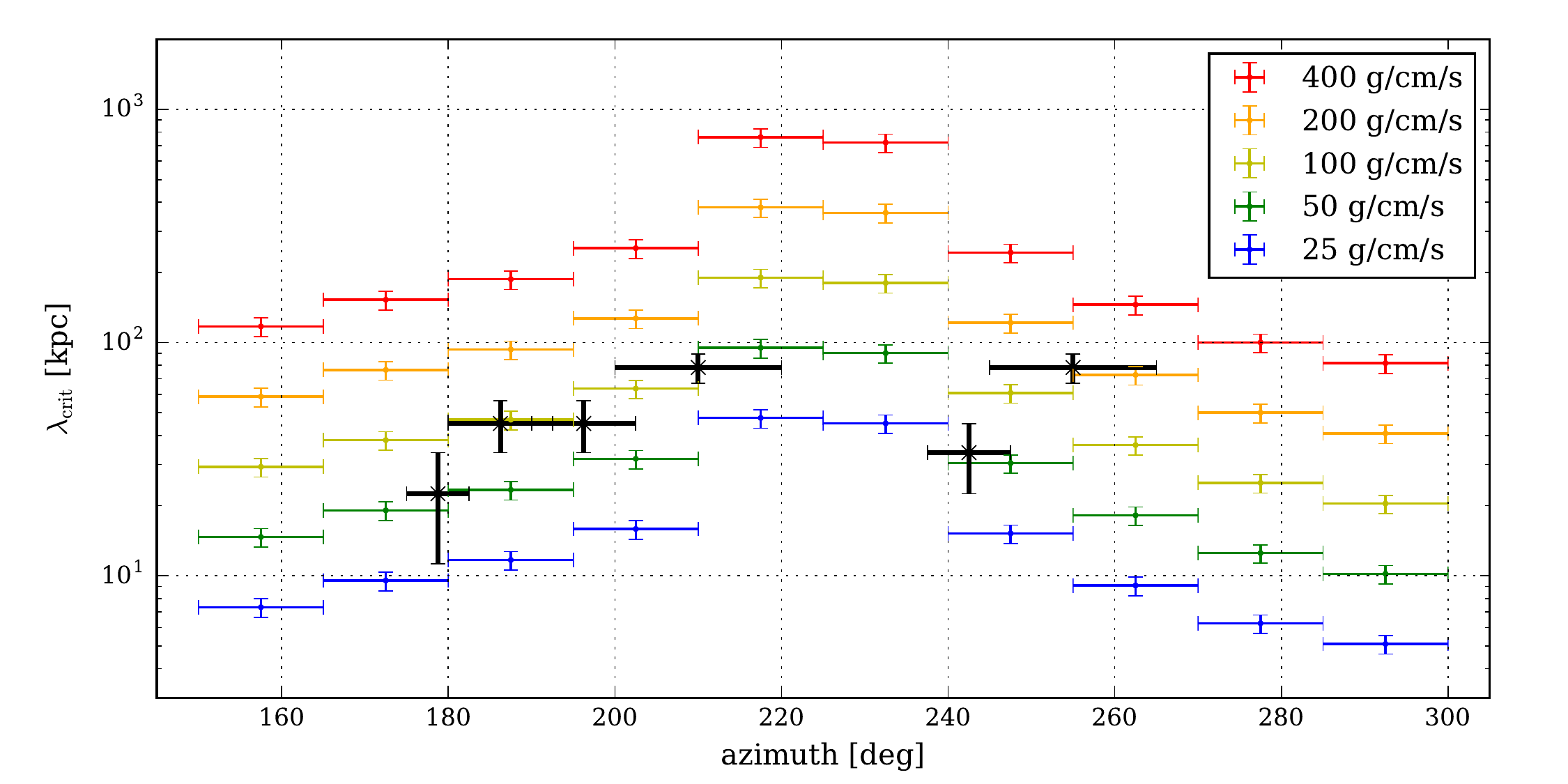}
 \caption[]{$\lambda_\mr{crit}$ values at each azimuth for various viscosity values. The instabilities whose length scales are smaller than $\lambda_\mr{crit}$ must be suppressed. The black crosses represent the sequential bumps that we pointed out in the previous section.}
 \label{img:visc}
\end{figure*}
Figure~\ref{img:visc} shows the $\lambda_\mr{crit}$ values at each azimuth for various viscosity values. In plotting them, we assumed that the axis of symmetry with respect to the ambient flow is at 225$^\circ$. For the density values, we used the deprojected ones (see Section~\ref{sec:deproj}). The black crosses roughly represent the sequential bumps that we pointed out in the previous section. Note that we neglected the effect of advection of the instabilities along the interface for simplicity.

The instabilities whose length scales are $<\lambda_\mr{crit}$ must be suppressed. However, in Figure~\ref{img:comp} we identified several instability candidates which we indicate as black crosses in Figure~\ref{img:visc}. Therefore, the viscosity above $\sim200~\mr{g/cm/s}$ is unlikely because we would not find the sequential bumps with such high viscosity values. Consequently, assuming the zeroth order analytical derivation of Equation~\ref{eq:visc}, the upper limit of the ICM viscosity is $\sim200~\mr{g/cm/s}$.

The estimated upper limit depends on the numerous assumptions that we made (e.g., geometry, KHI scenario and existence and scales of the bumps). In addition, it is difficult to break it down into the many kinds of physical processes which may affect the effective value. For example, the perturbations on smaller scales are suppressed not only by viscosity, but also by e.g. the surface tension at the interface or the finite width of the interface \citep{landau,churazov04b,roediger13b}. The magnetic field which inevitably exists also plays a role; it might be acting as a surface tension term, or the effect of anisotropy of the viscosity (Braginskii viscosity) may be non-negligible \citep{zuhone15}.

To interpret our estimated upper limit of the effective viscosity of the ICM, we compare the value of $\mu\lesssim 200~\mr{g/cm/s}$ to the fiducial viscosity of plasma, expressed as the isotropic Spitzer-like temperature-dependent viscosity \citep{spitzer,sarazin86,roediger13a}
\begin{equation}
 \mu = 5200~\mr{g/cm/s}\left(\frac{kT}{8.4~\mr{keV}}\right)^{5/2},
\end{equation}
where 8.4~keV is the typical temperature just outside the interface and the Coulomb logarithm $\ln\Lambda = 40$ is assumed. Our estimated value is suppressed to at most 5\% of the isotropic Spitzer-like temperature-dependent viscosity. This suppression is likely due to the magnetic field lines parallel to the interface which freeze in the ambient flow along the interface \citep[magnetic draping;][]{lyutikov06,dursi08,ruszkowski14}. Because of the small gyroradii, momentum transfer across the field lines should be reduced. Note that the existence of such a layer of magnetic field lines is consistent with the widths of the interface that are thinner than the Coulomb mean free paths \citep{vikhlinin01b,vikhlinin02,datta14}.

The suppression of the effective viscosity from the Spitzer value at the interface of cold fronts (previous paragraph) has indeed been suggested by several observations and numerical simulations \citep[e.g.][]{werner15,su16,roediger13a,roediger13b}. Moreover, interestingly, the corresponding upper limit of the kinematic viscosity $\nu\lesssim\mu/\rho\sim1.4\times 10^{29}~\mr{cm^2/s}$ is consistent with the previous upper limit $\nu < 3\times 10^{29}~\mr{cm^2/s}$ \citep[Coma cluster,][]{schuecker04} and the lower limit $\nu > 4\times 10^{27}~\mr{cm^2/s}$ \citep[Perseus cluster,][]{fabian03}, which are estimated for different targets and in different methods.

Generally, the ICM viscosity should depend on other physical parameters such as the gas temperature and the strength and configuration of the magnetic field. Therefore, it is not necessarily expected that the ICM viscosity has a universal value. On the other hand, the above mentioned facts may imply the intriguing universality of the effective kinematic viscosity of the ICM.

\subsection{Possible vortex street}
We point out the brightness dip and excess in the X-ray image (see the black arrows in Figure~\ref{img:flatimage}). The morphology of this feature is similar to that of the von K\'{a}rm\'{a}n vortex.

By combining the limits on the kinematic viscosity presented in the last section, we can assume the kinematic viscosity of the ICM in this system has the value of $\nu\sim10^{28-29}~\mr{cm^2/s}$, in which case the corresponding viscosity $\mu\sim100~\mr{g/cm/s}$. Using this value, the Reynolds number corresponding to the overall morphology whose spatial scale is $\sim500~\mr{kpc}$ and assuming a speed of $950~\mr{km/s}$ is estimated at $\mr{Re}\sim2100$. In the case of a sphere, the hairpin vortex is established if $\sim300<\mr{Re}<3.7\times10^5$, and if $650 < \mr{Re}$ the vortex street becomes turbulent \citep{kiya01}. Thus, it is possible that the hairpin vortex has been established in this system.

The vortex shedding frequency $f=SV/\lambda$, where $S$ is the Strouhal number and $S\sim0.2$ for the hairpin vortex regime, is calculated at $f\sim0.4~\mr{Gyr^{-1}}$. Since the typical dynamical time-scale of galaxy clusters is of order Gyr, there must not be multiple vortices, which is consistent with our observation. Note that, if we use the Spitzer viscosity in the calculation above, it does not yield the Reynolds number within the hairpin vortex regime ($\mathrm{Re}\sim40$).

\section{Conclusions}
In this paper, we studied the cold front in Abell~3667 using $\sim$500~ksec archival {\it Chandra} data. The main results of our work are summarized below.
 \begin{enumerate}
  \item We find a striking similarity between the cold front in Abell~3667 and recent numerical simulations of inviscid gas-stripping. This indicates that the front is formed by a merger which takes place nearly in the plane of the sky and the effective viscosity of the ICM is suppressed.
  \item By extracting azimuthally resolved surface brightness profiles, we find that the radii of the interface fluctuate with position angle. The characteristic length scale of the fluctuations is 20--80~kpc.
  \item We find multiple edges in the surface brightness profiles along some directions. The ratio of the distance of the edge to the typical scale of the fluctuation in the azimuthal direction is consistent with the prediction of numerical simulations of cold fronts. In one direction, we find a dip in the {\it projected} pressure profile, which we interpret as a KHI layer.
  \item We propose a scenario where KHIs are developing along the cold front in Abell 3667. This is the first observational indication of developing KHIs along a merger cold front in a galaxy cluster.
  \item Based on the assumption that KHIs are actually developing along the interface, we estimated the upper limit of the ICM effective viscosity. The estimated upper limit of $\mu\lesssim200~\mr{g/cm/s}$ is at most 5\% of the isotropic Spitzer-like temperature dependent viscosity, as has been suggested by several observations and numerical simulations. The corresponding upper limit of the kinematic viscosity $\nu\lesssim1.4\times10^{29}~\mr{cm^2/s}$ is consistent with the previous upper and lower limits which were determined for other targets and by different methods.
  \item The temperature and the entropy jumps are relatively moderate at the edge of the front compared with the tip of the front. This difference is probably due to the mixing of the gas induced by the fully developed, turbulent KHIs.
 \end{enumerate}

\section*{Acknowledgements}
YI is financially supported by a Grant-in-Aid for Japan Society for the Promotion of Science (JSPS) Fellows (16J02333). This project has been supported by the Lend\"{u}let LP2016-11 grant awarded by the Hungarian Academy of Sciences. We thank Dr. Elke Roediger and IOP Publishing for the permission of reproducing their figures in our article.





\bibliographystyle{mnras}
\bibliography{ref}



\appendix
\section{Thermodynamic mapping}\label{appendix:thermo}
\begin{figure*}
 \begin{minipage}{0.495\hsize}
  \centering
  \includegraphics[width=3.0in]{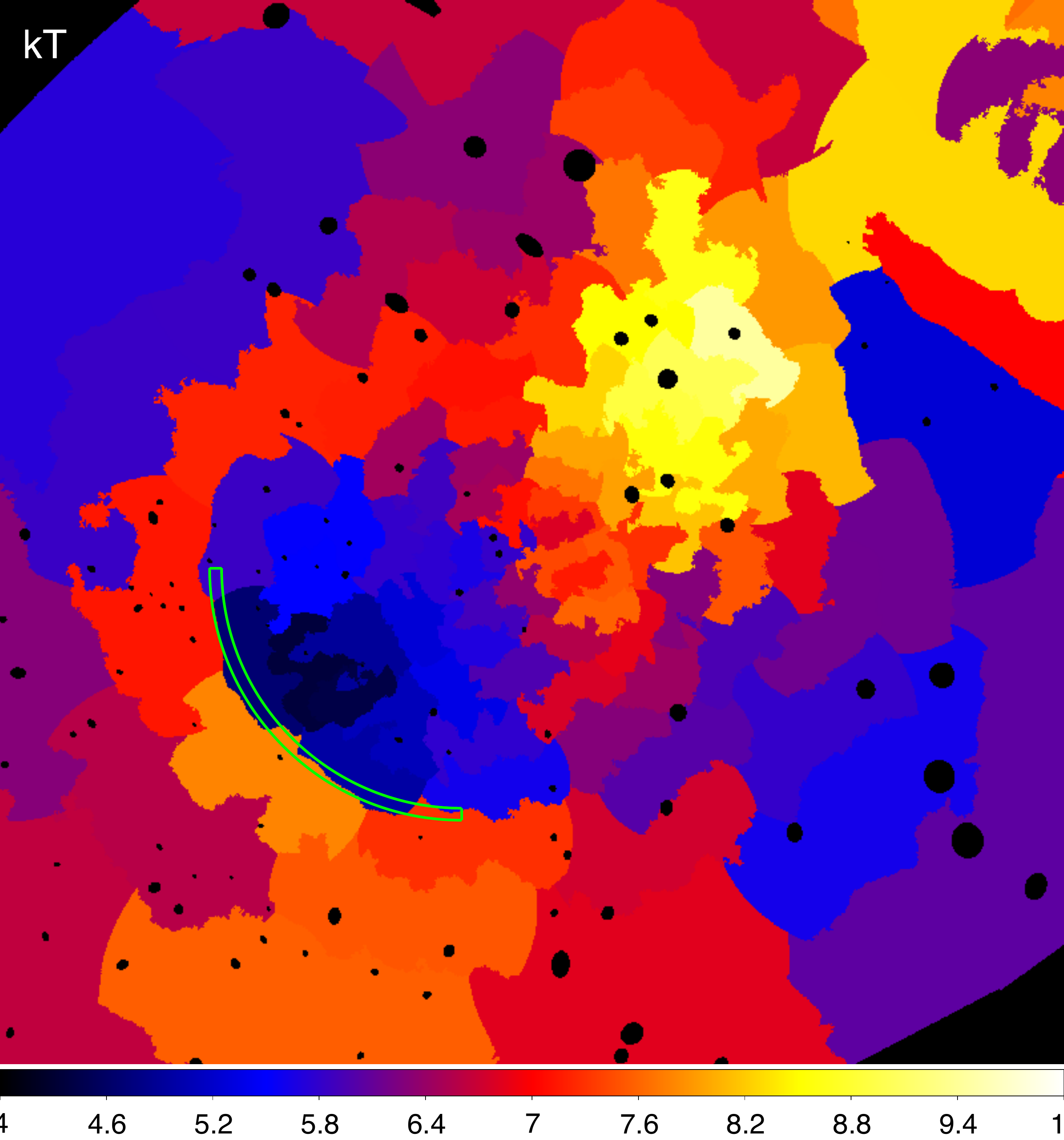}
 \end{minipage}
 \begin{minipage}{0.495\hsize}
  \centering
  \includegraphics[width=3.0in]{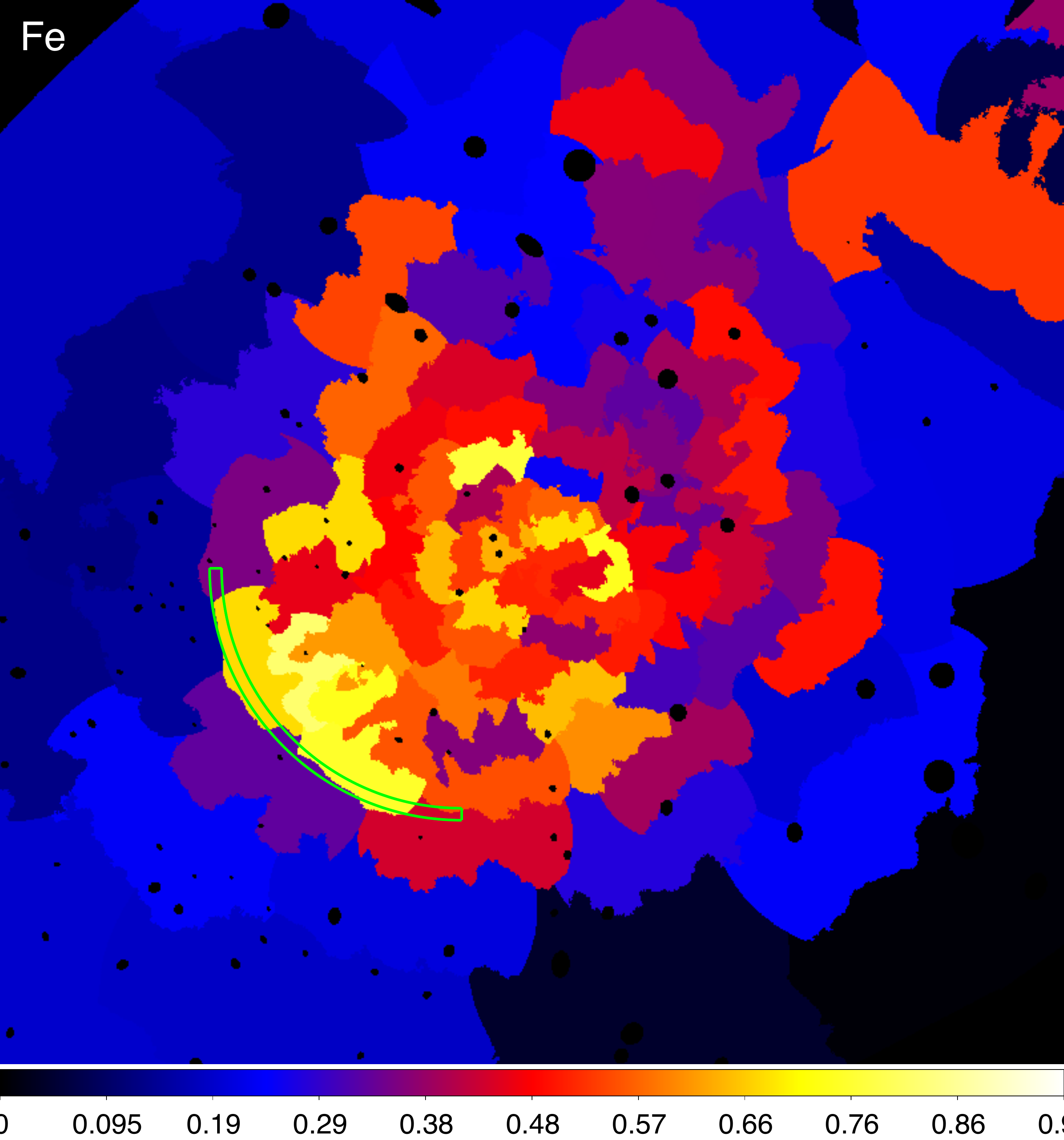}
 \end{minipage}
 \begin{minipage}{0.495\hsize}
  \centering
  \includegraphics[width=3.0in]{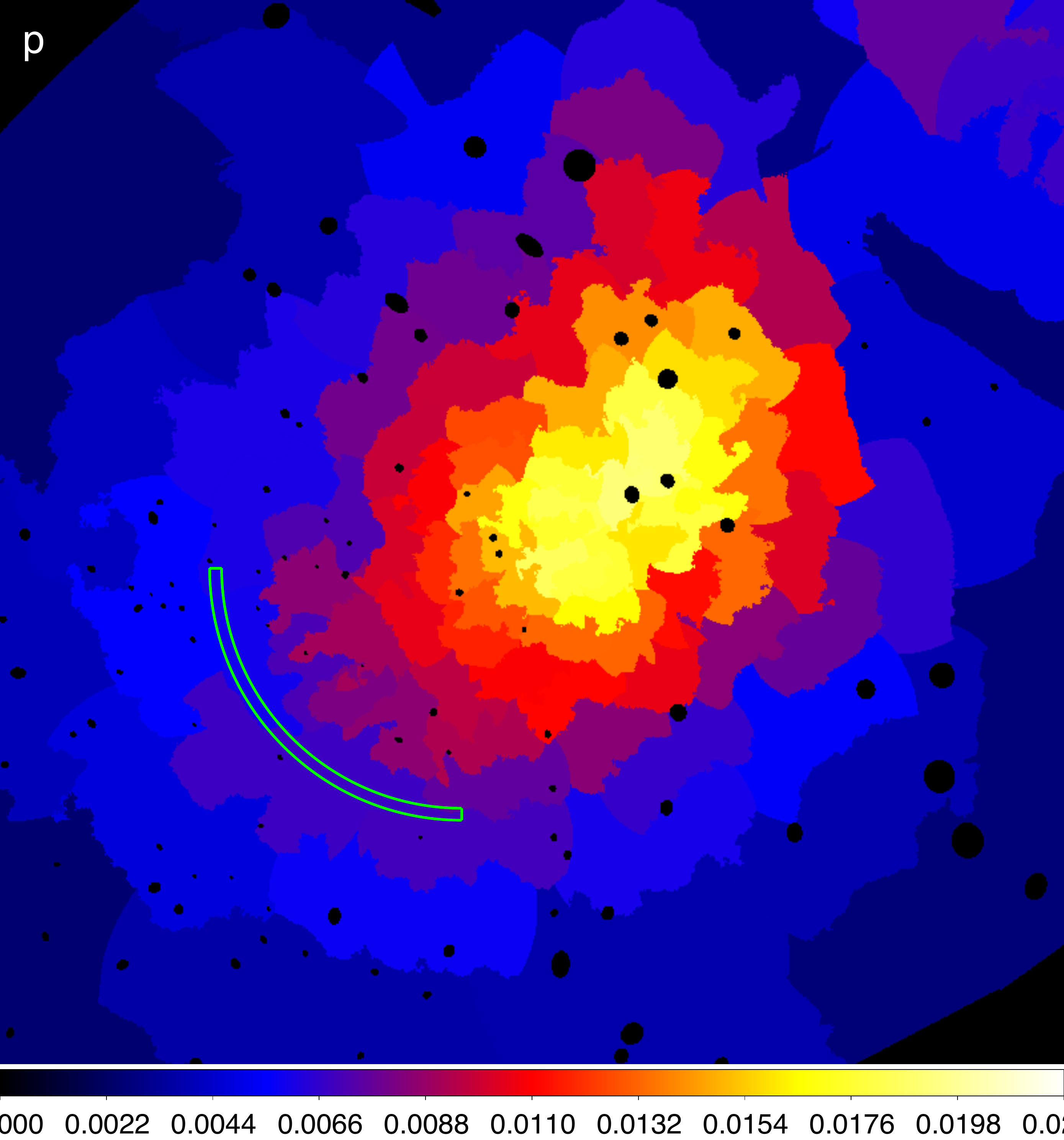}
 \end{minipage}
 \begin{minipage}{0.495\hsize}
  \centering
  \includegraphics[width=3.0in]{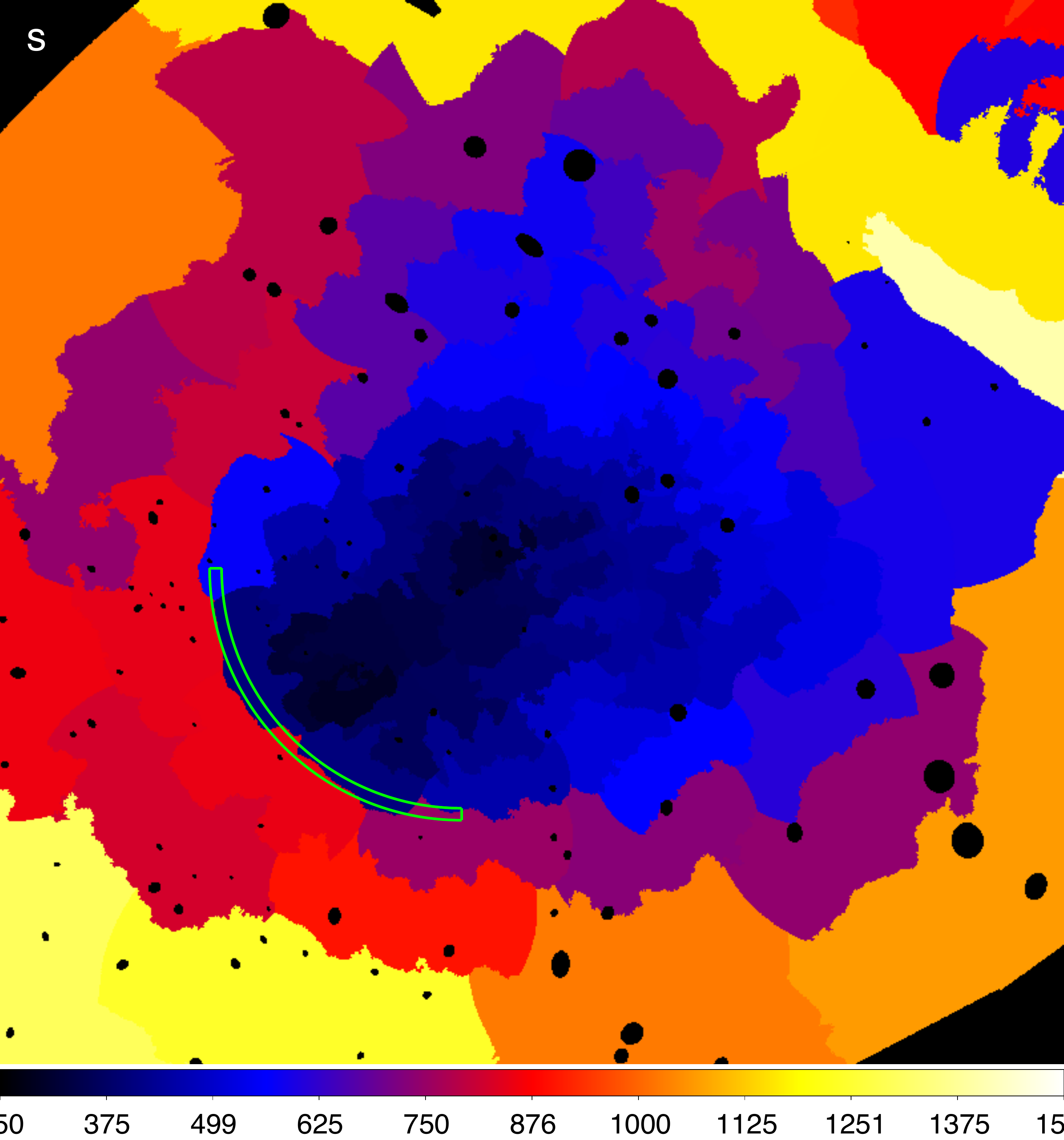}
 \end{minipage}
 \caption[]{{\it Projected} thermodynamic maps assuming a uniform ICM line-of-sight depth of $l=1$~Mpc. {\it Top left:} {\it projected} temperature map in the unit of keV. {\it Top right:} {\it projected} Fe abundance map with respect to the solar abundance by \citet{lodders03}. {\it Bottom left:} {\it projected} electron pressure map in the unit of $\mr{keV cm}^{-3}\times(l/1~\mr{Mpc})^{-1/2}$. {\it Bottom right:} {\it projected} entropy map in the unit of $\mr{keV cm}^2\times(l/1~\mr{Mpc})^{1/3}$. The location of the cold front is denoted in green.}
 \label{img:thermo}
\end{figure*}

To explore the two-dimensional thermodynamic structure, we conducted ({\it projected}) thermodynamic mapping. We adopted the contour binning algorithm \citep{sanders06}, which divides the entire field of view into smaller regions, based on the statistical error of each bin calculated from the raw and the background counts, so that all the bins have similar statistical signal-to-noise (S/N) ratio. The S/N ratio of each bin is about 150, corresponding to about 23000~counts/bin.

For each of the bins, we extracted the spectra and fit them using a single-temperature thermal plasma model absorbed by the Galactic absorption (i.e. assuming that the spectra are represented by single-temperature thermal plasma). For the spectral fitting, we used {\small XSPEC} (version 12.8.2) \citep{arnaud96} to minimize $\chi^2$. We used the chemical abundance table determined by \citet{lodders03}.

We translated the best-fitting normalization value to the physical number density of the ICM, assuming that the ICM is uniform along the line-of-sight, that the line-of-sight depth of the ICM is also uniform over the entire field of view with the value of $l=1$~Mpc, and that the electron density $n_{\rm e}$ and hydrogen ion density $n_H$ satisfy a relation $n_{\rm e} = 1.2n_H$. The calculated electron density $n_{\rm e}$ is combined with the best-fitting temperature value $kT$ for the calculation of the electron pressure ($p=n_{\rm e}kT$) and the entropy values ($s=kTn_{\rm e}^{-2/3}$).

 The resulting temperature, Fe abundance, electron pressure and entropy maps are shown in Figure~\ref{img:thermo}. The black ellipses are the positions of point sources which are detected using the \verb+wavdetect+ tool in the {\small CIAO} software package with the scales of 1, 2, 4, 8, 16~pixels.

 \section{Results of surface brightness profile fitting}\label{appendix:sbfit}
 \begin{figure*}
 \begin{minipage}{0.333\hsize}
  \centering
  \includegraphics[width=2.3in]{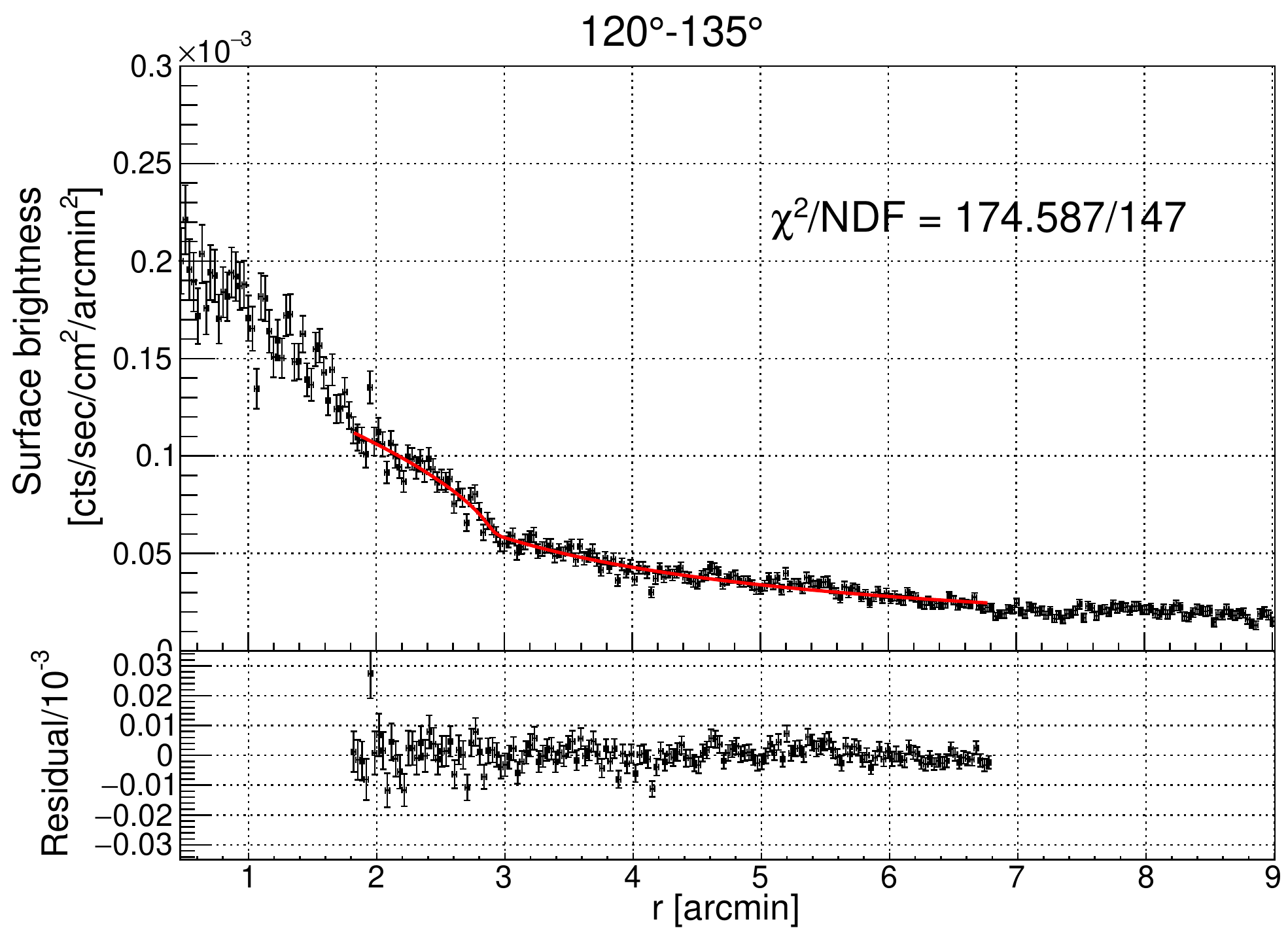}
 \end{minipage}%
 \begin{minipage}{0.333\hsize}
  \centering
  \includegraphics[width=2.3in]{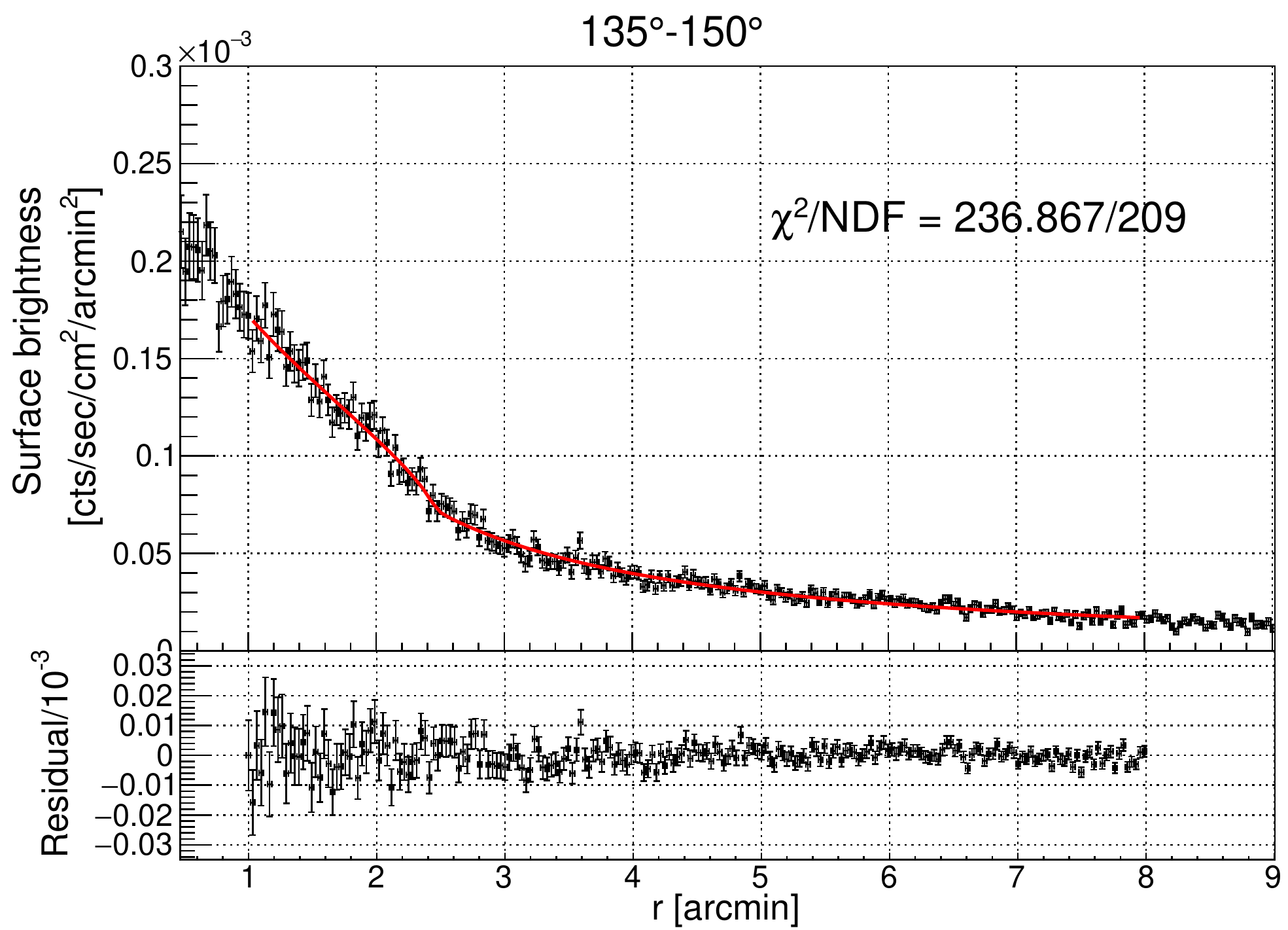}
 \end{minipage}%
 \begin{minipage}{0.333\hsize}
  \centering
  \includegraphics[width=2.3in]{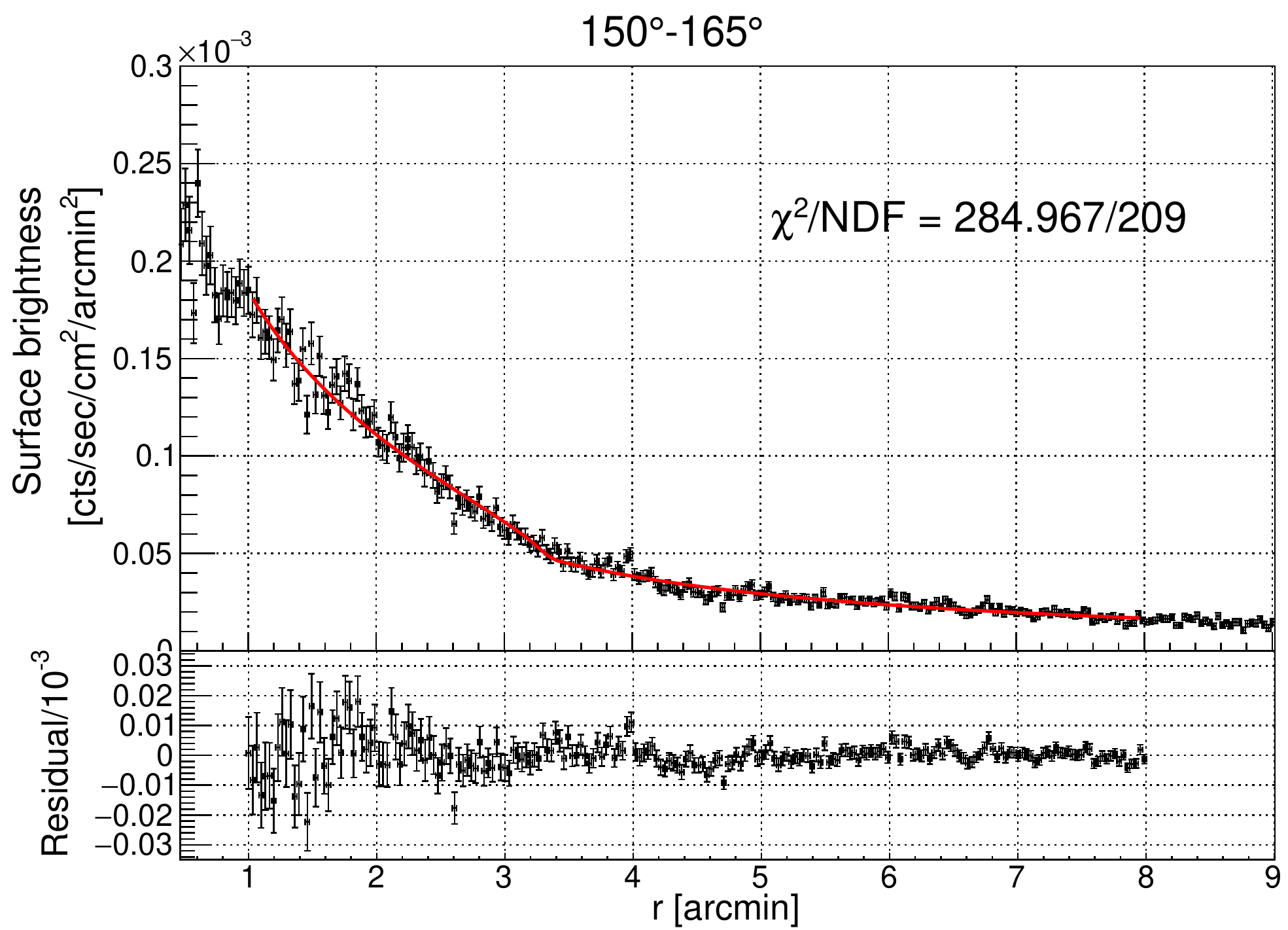}
 \end{minipage}
 \begin{minipage}{0.333\hsize}
  \centering
  \includegraphics[width=2.3in]{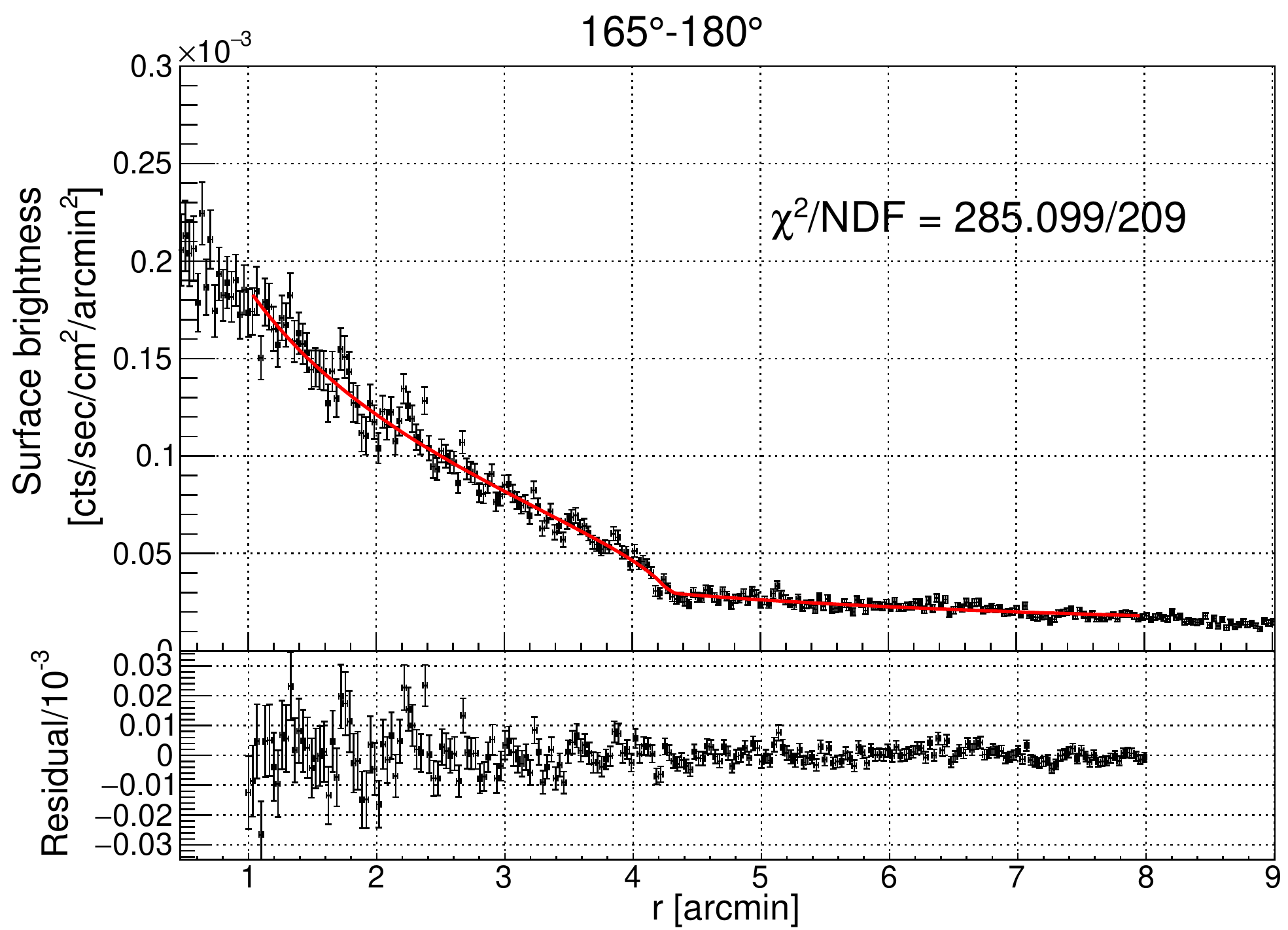}
 \end{minipage}%
 \begin{minipage}{0.333\hsize}
  \centering
  \includegraphics[width=2.3in]{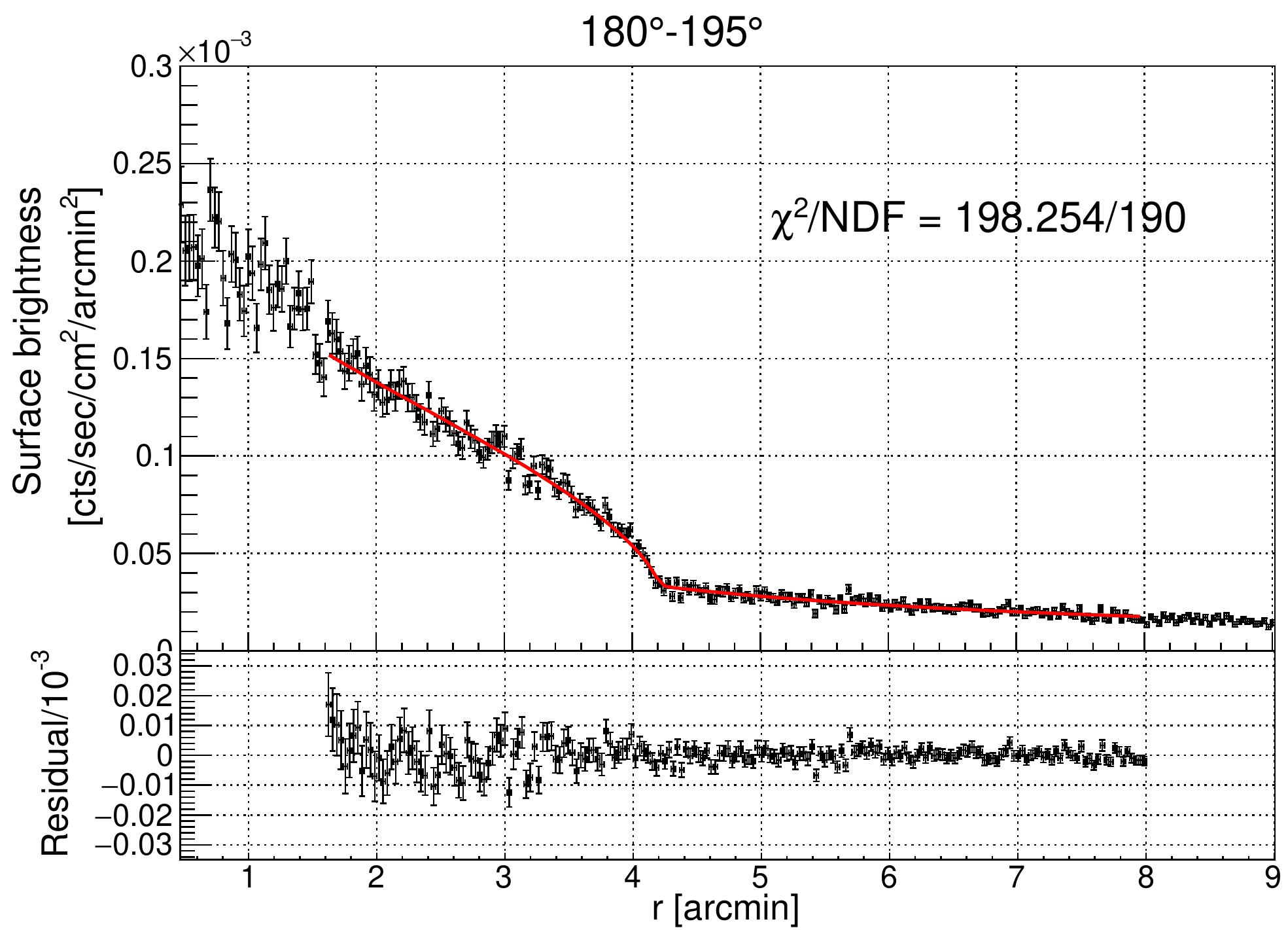}
 \end{minipage}%
 \begin{minipage}{0.333\hsize}
  \centering
  \includegraphics[width=2.3in]{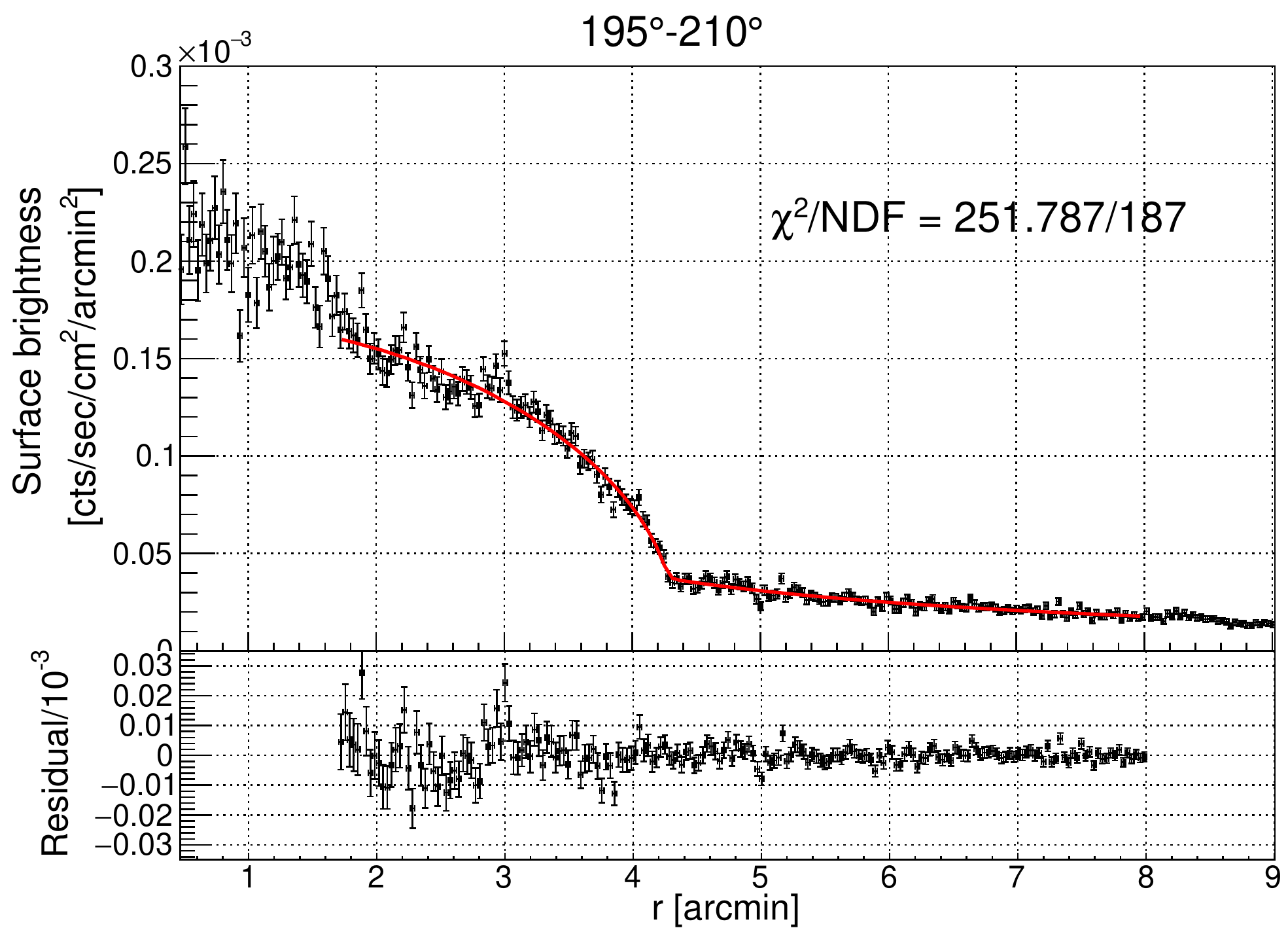}
 \end{minipage}
 \begin{minipage}{0.333\hsize}
  \centering
  \includegraphics[width=2.3in]{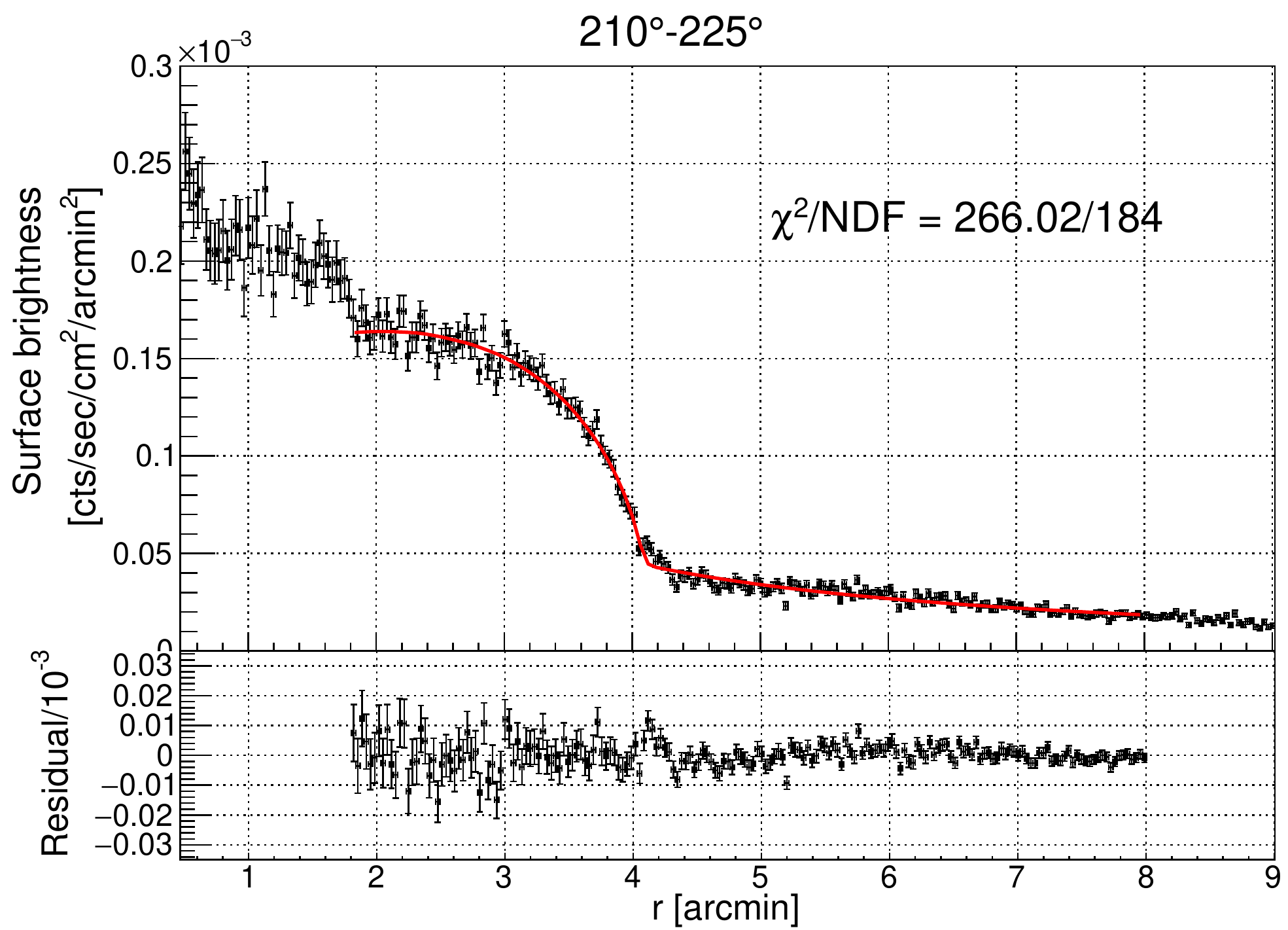}
 \end{minipage}%
 \begin{minipage}{0.333\hsize}
  \centering
  \includegraphics[width=2.3in]{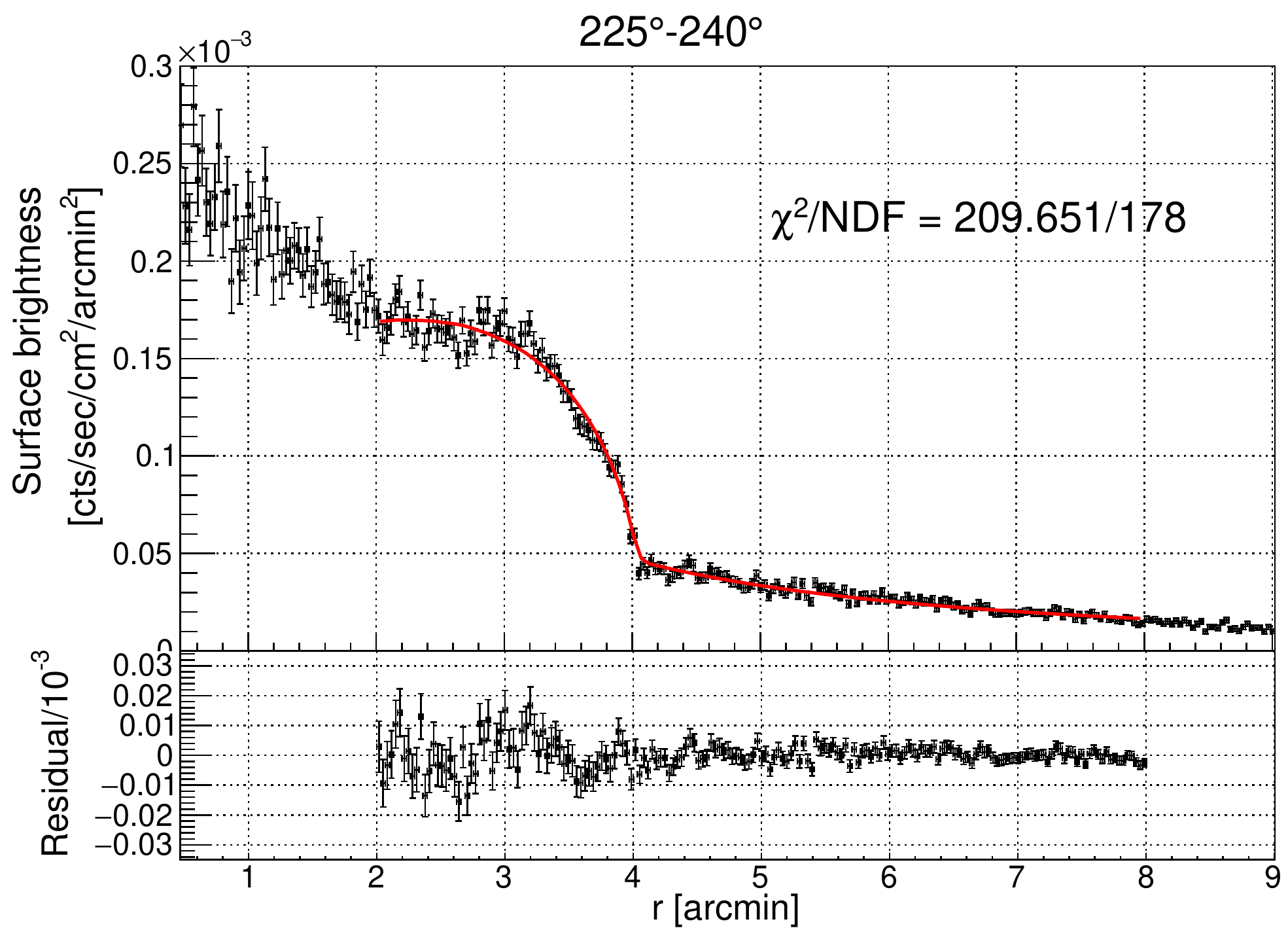}
 \end{minipage}%
 \begin{minipage}{0.333\hsize}
  \centering
  \includegraphics[width=2.3in]{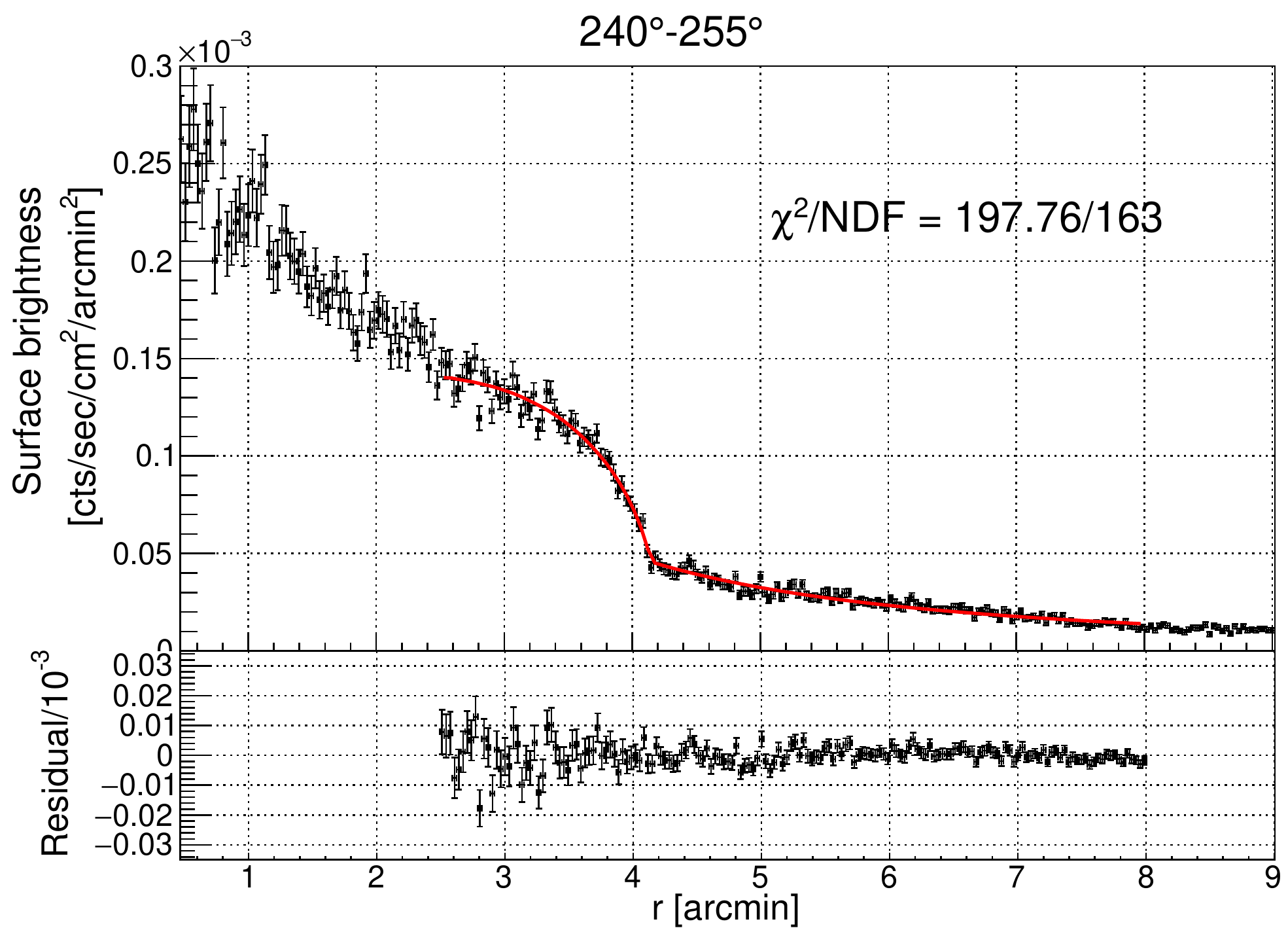}
 \end{minipage}
 \begin{minipage}{0.333\hsize}
  \centering
  \includegraphics[width=2.3in]{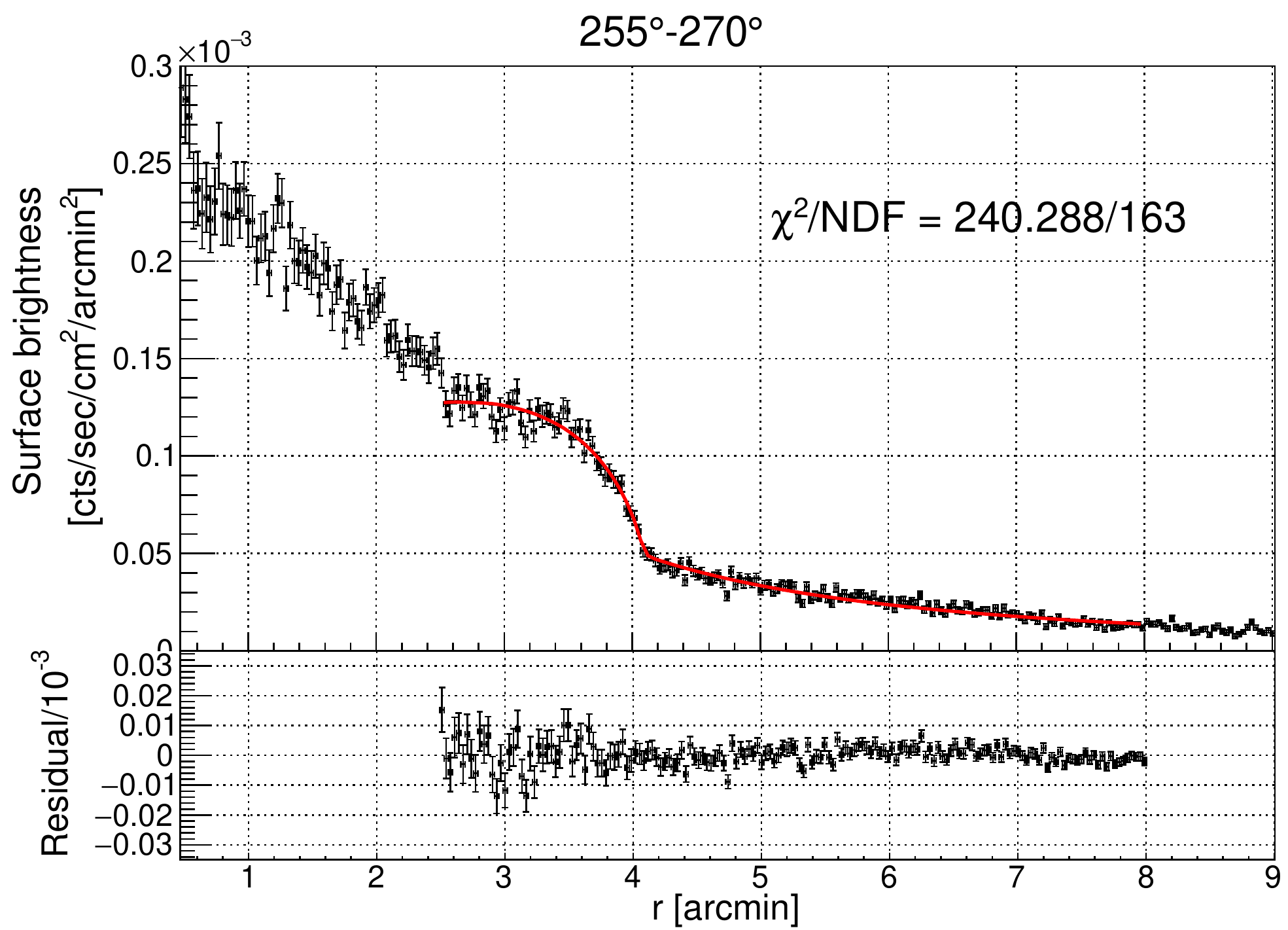}
 \end{minipage}%
 \begin{minipage}{0.333\hsize}
  \centering
  \includegraphics[width=2.3in]{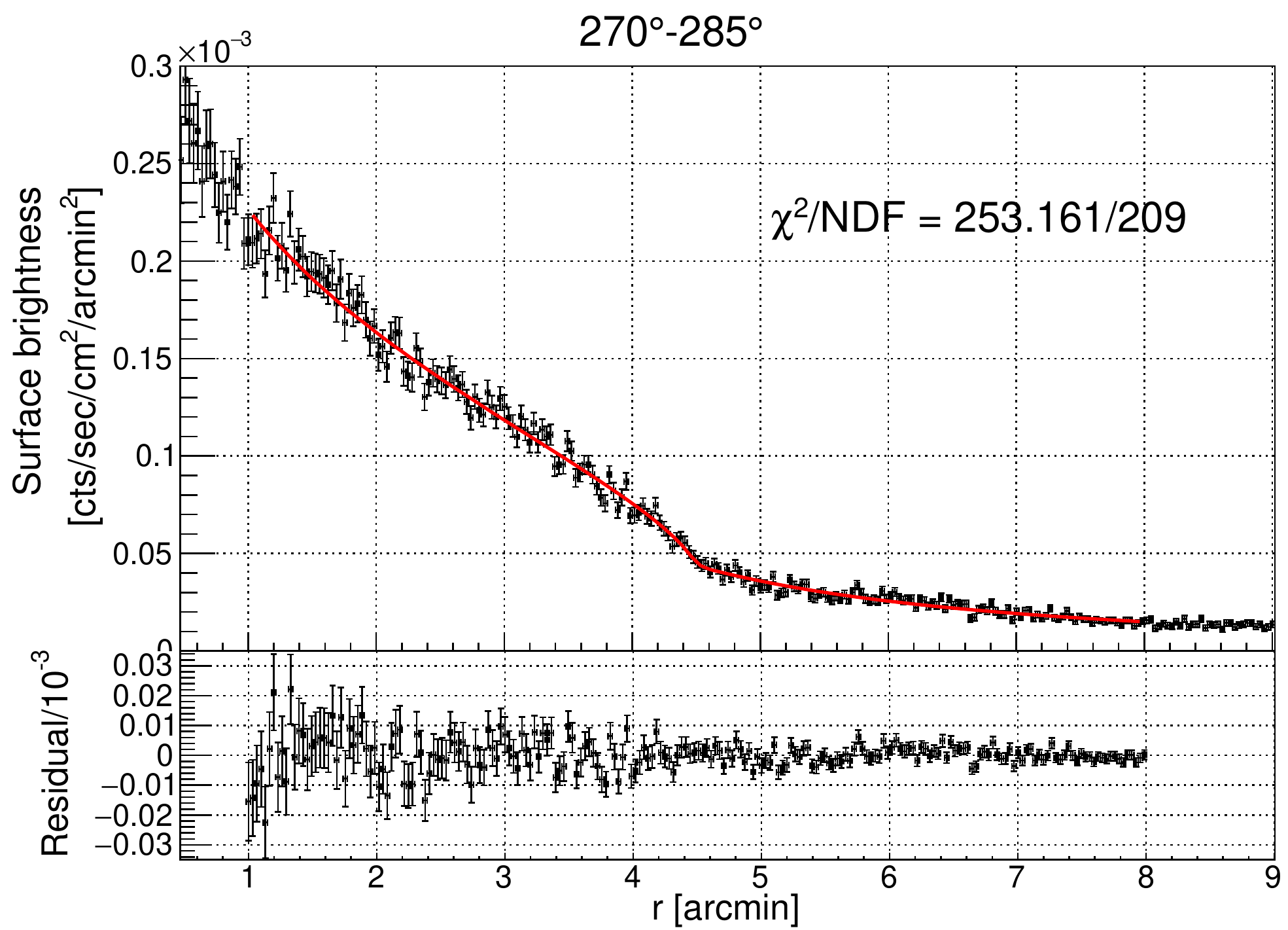}
 \end{minipage}%
 \begin{minipage}{0.333\hsize}
  \centering
  \includegraphics[width=2.3in]{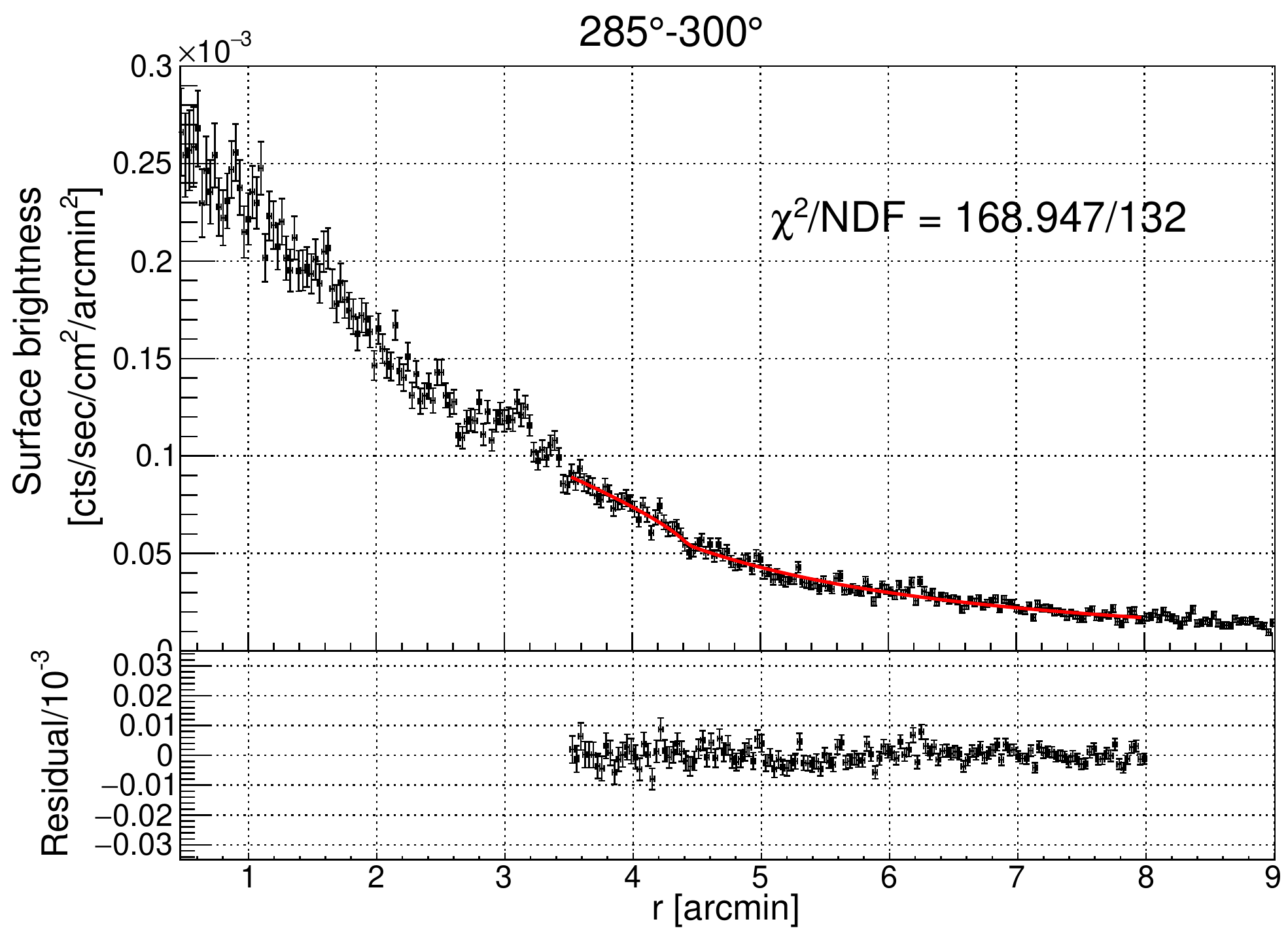}
 \end{minipage}
 \begin{minipage}{0.333\hsize}
  \centering
  \includegraphics[width=2.3in]{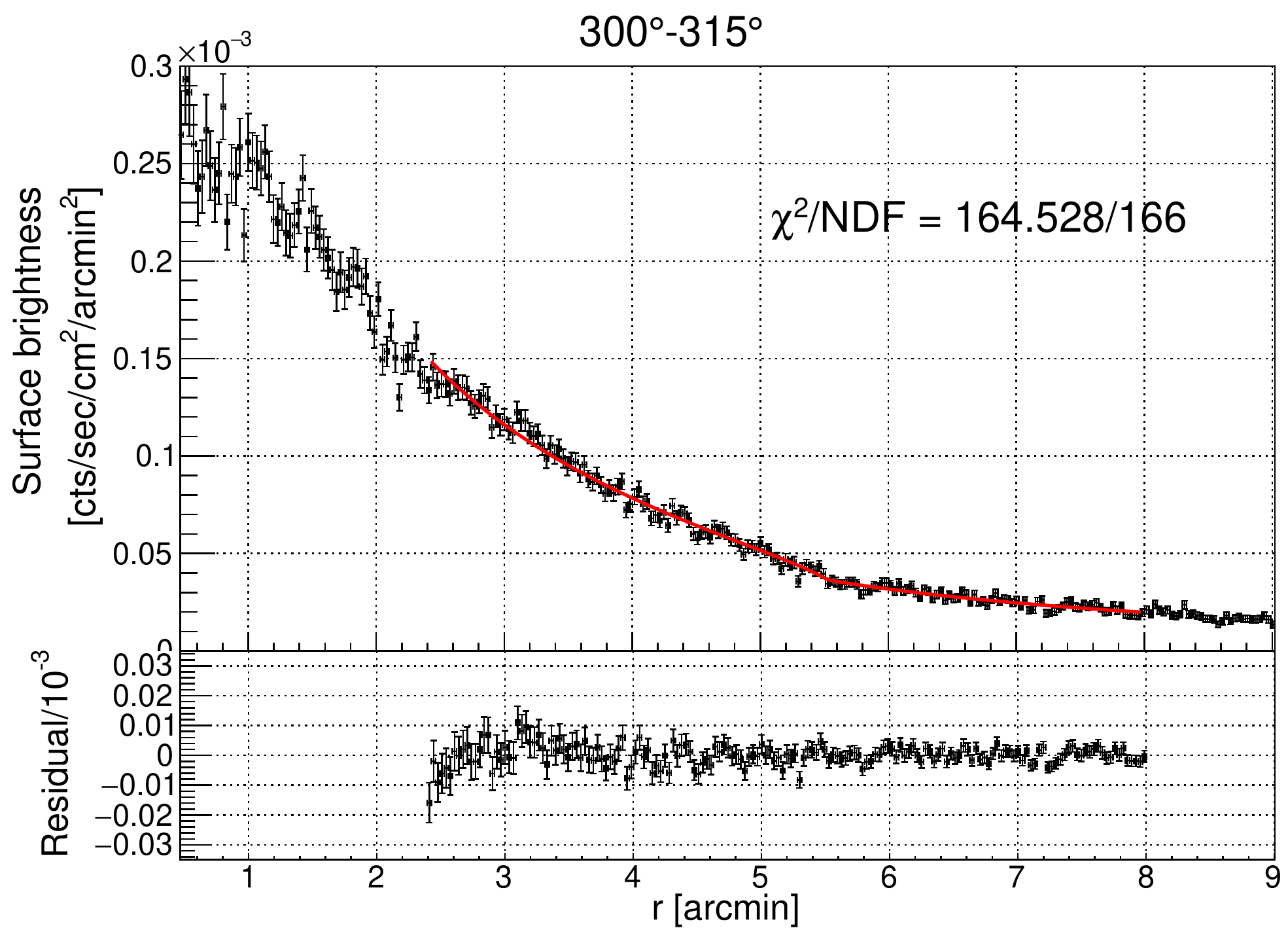}
 \end{minipage}%
 \begin{minipage}{0.333\hsize}
  \centering
  \includegraphics[width=2.3in]{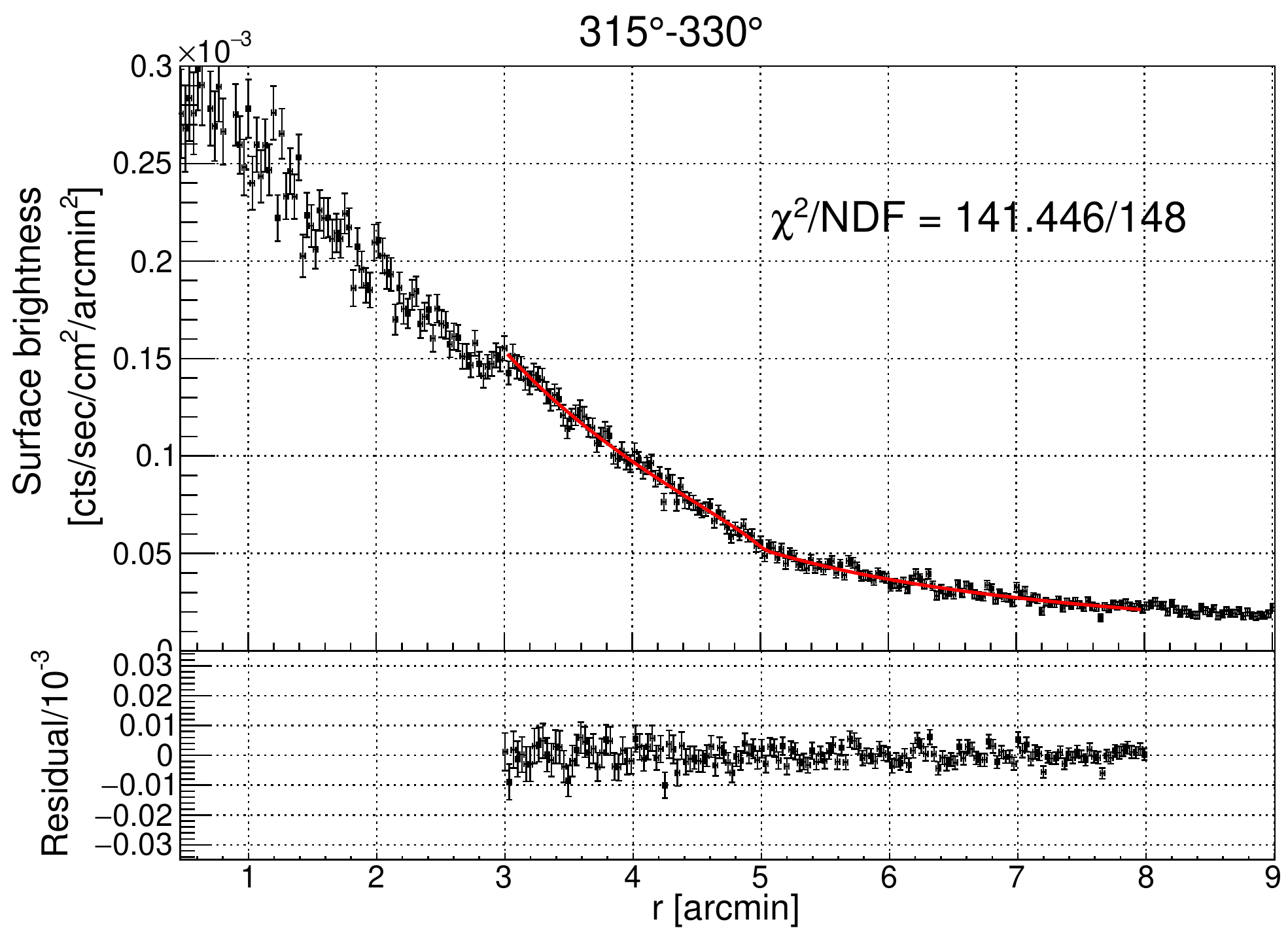}
 \end{minipage}%
 \caption[]{The surface brightness profiles extracted using 15$^\circ$ sectors. The red curves are the best-fitting projected broken power law models.}
\label{img:prof_sb}
 \end{figure*}

\section{Deprojected thermodynamic profiles}
\begin{figure*}
 \begin{minipage}{0.333\hsize}
  \centering
  \includegraphics[width=2.3in]{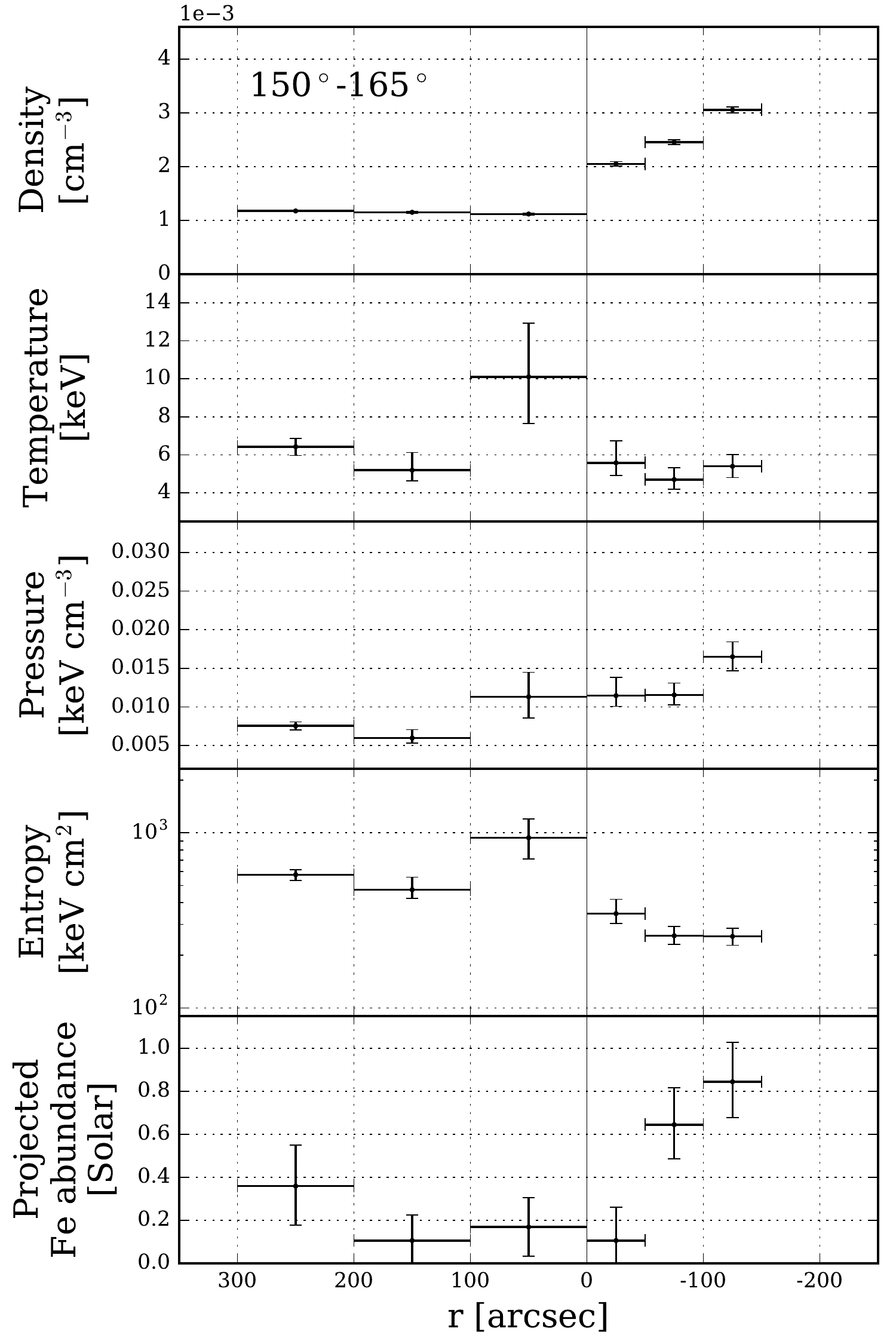}
 \end{minipage}%
 \begin{minipage}{0.333\hsize}
  \centering
  \includegraphics[width=2.3in]{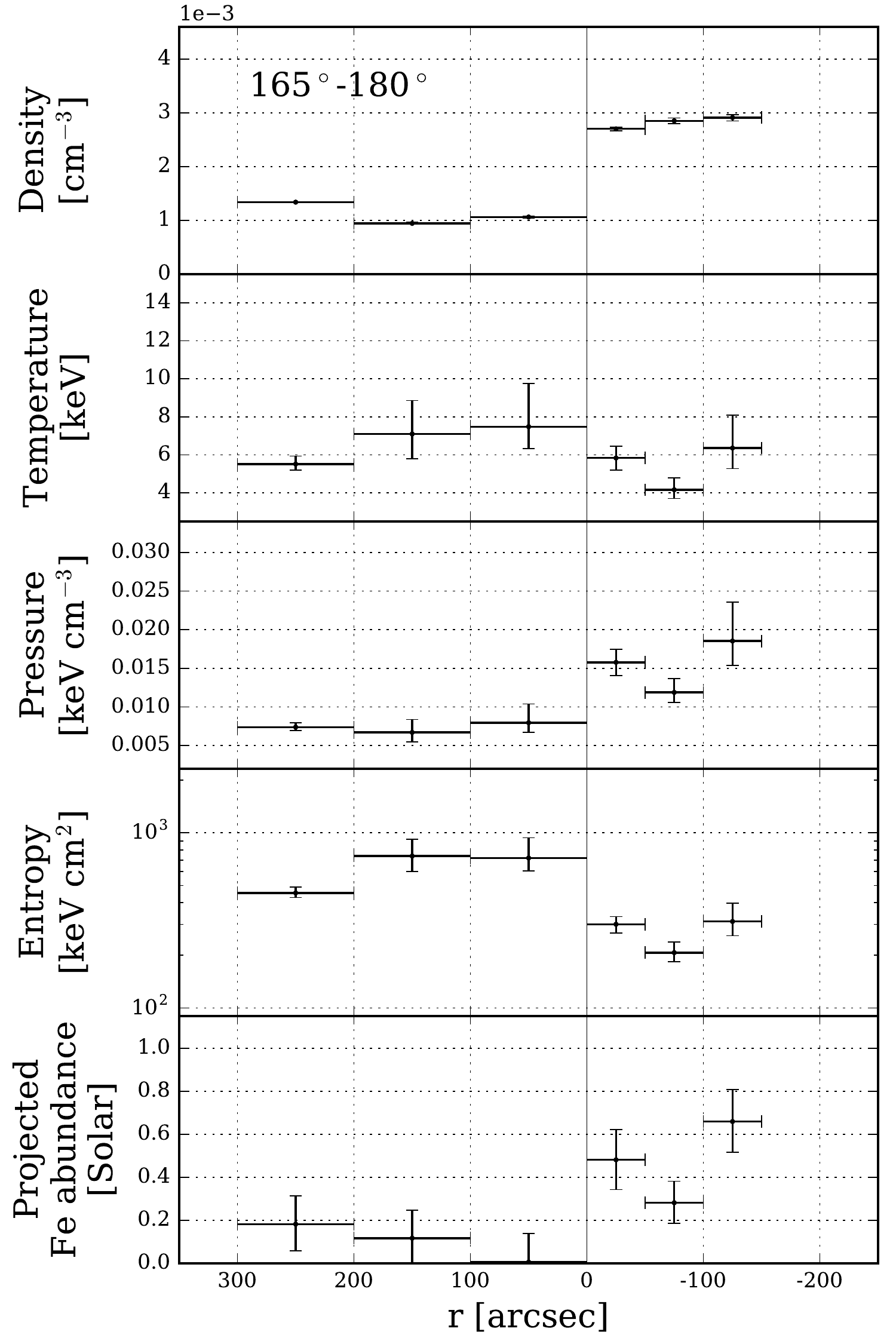}
 \end{minipage}%
 \begin{minipage}{0.333\hsize}
  \centering
  \includegraphics[width=2.3in]{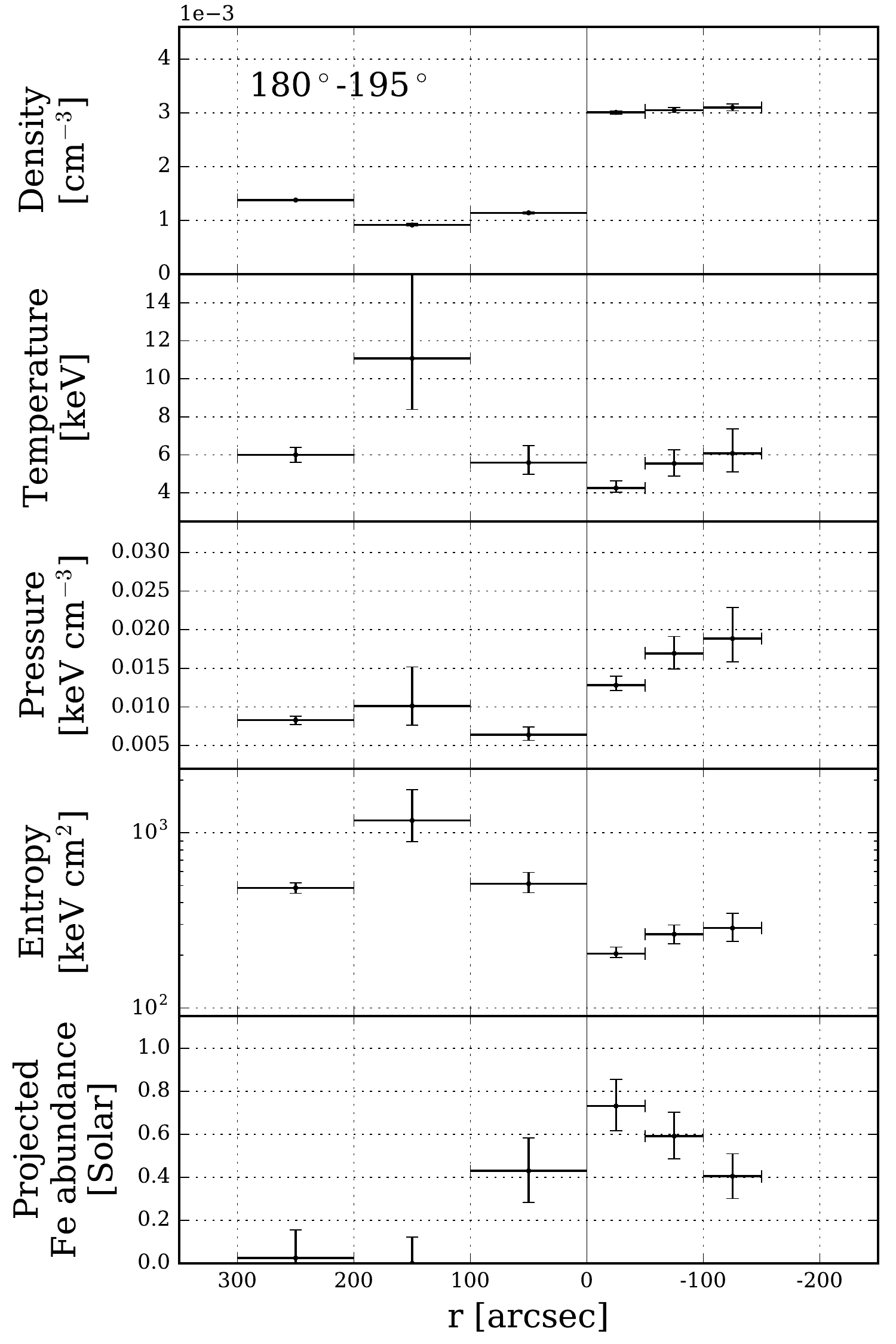}
 \end{minipage}
 \begin{minipage}{0.333\hsize}
  \centering
  \includegraphics[width=2.3in]{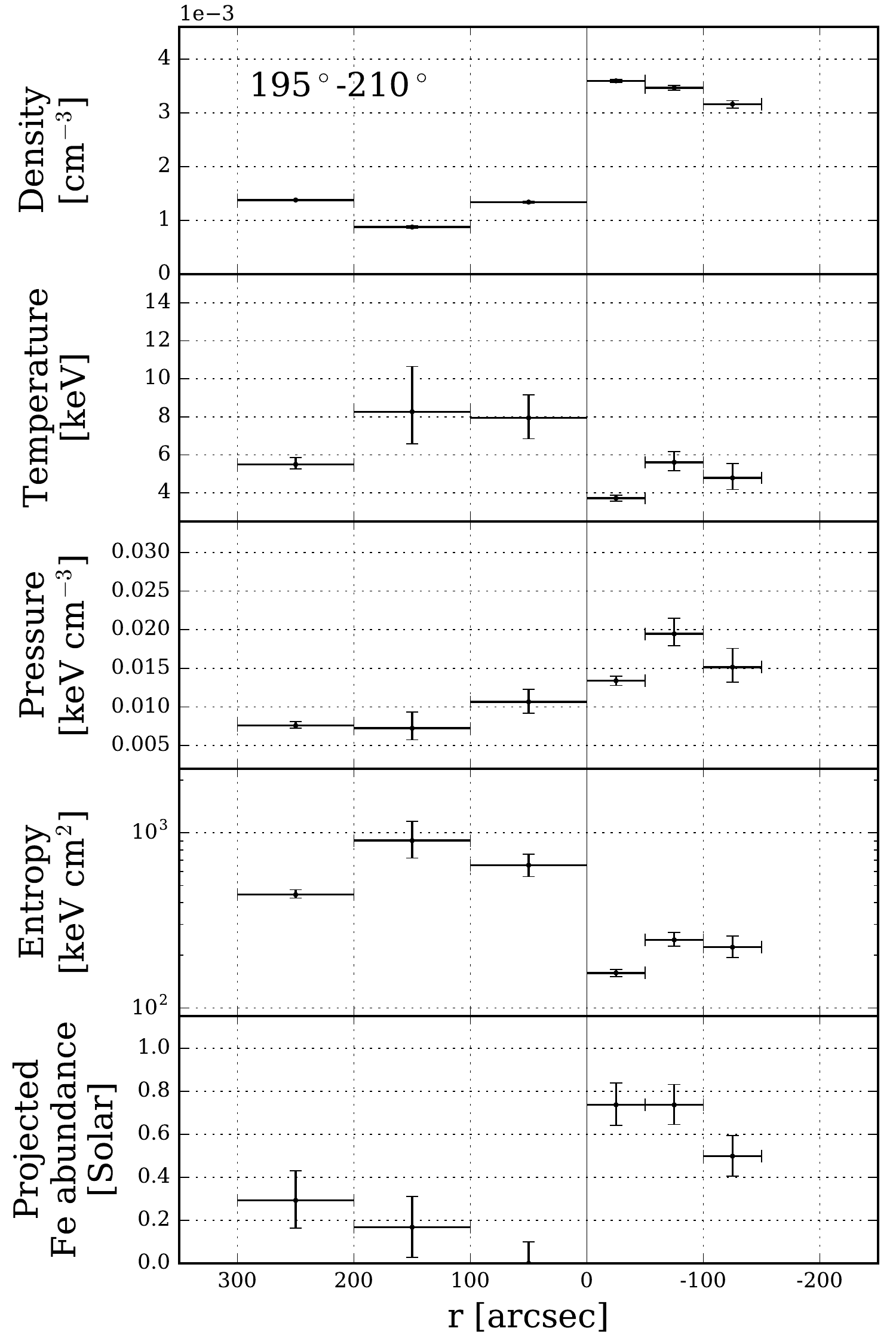}
 \end{minipage}%
 \begin{minipage}{0.333\hsize}
  \centering
  \includegraphics[width=2.3in]{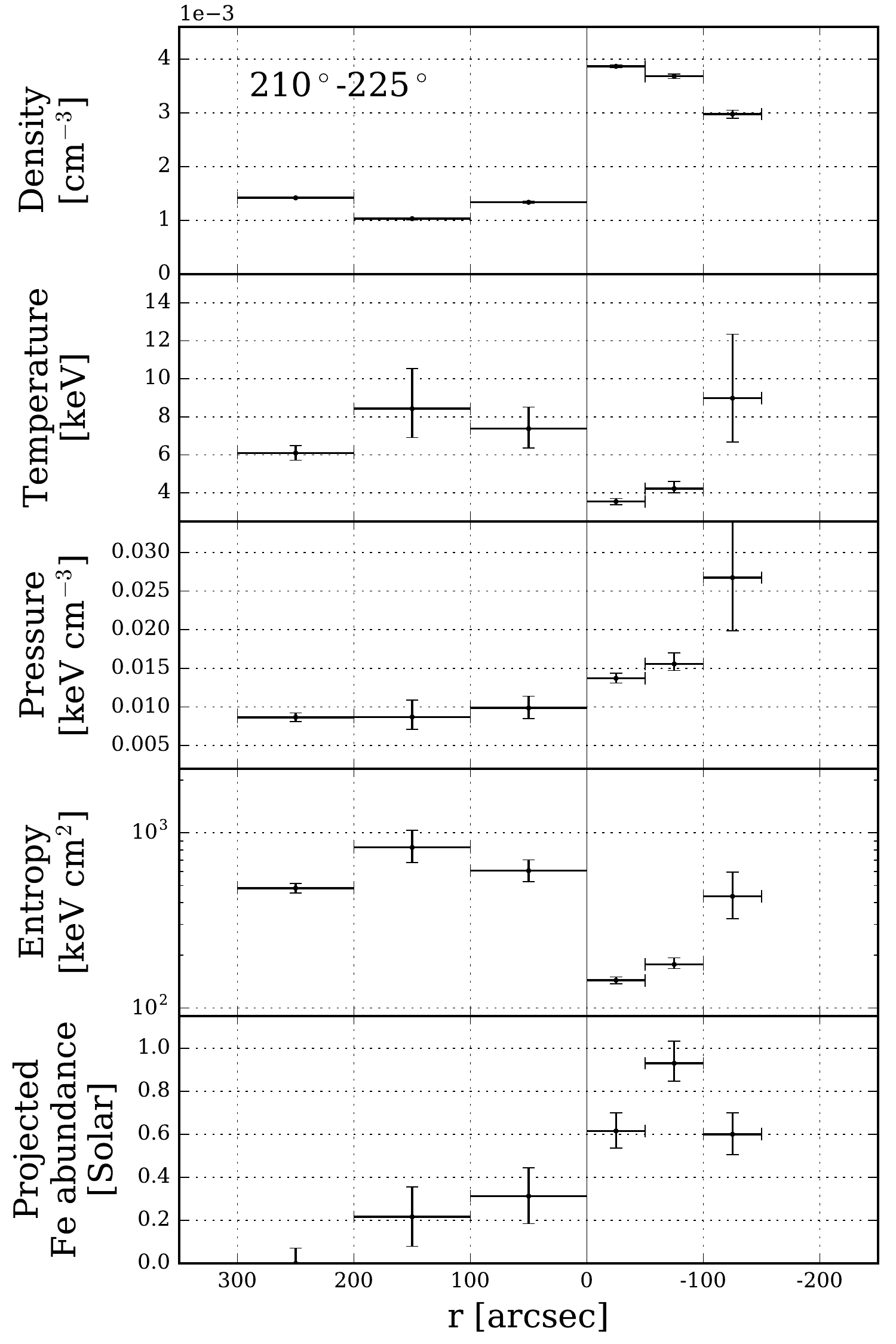}
 \end{minipage}%
 \begin{minipage}{0.333\hsize}
  \centering
  \includegraphics[width=2.3in]{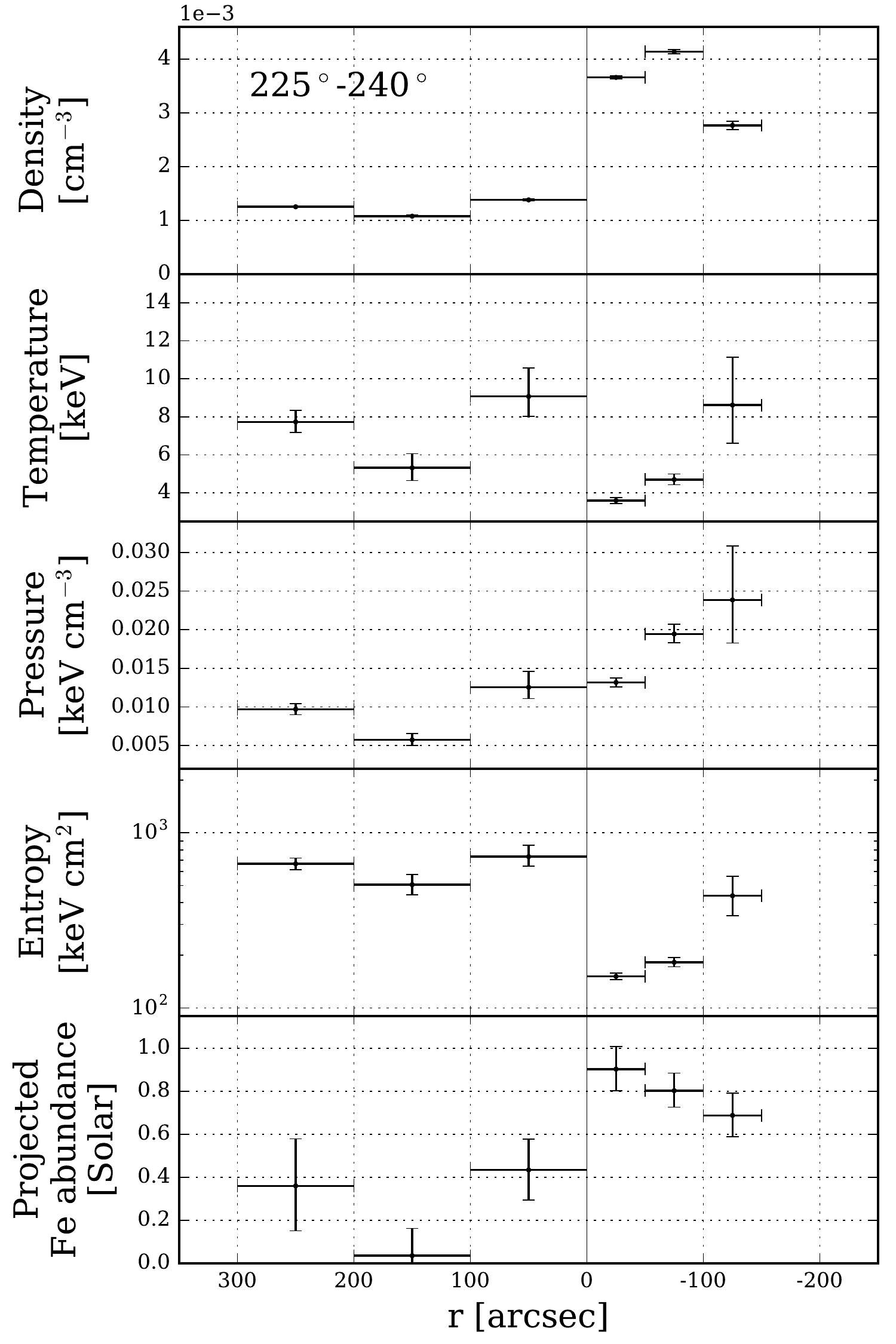}
 \end{minipage}
 \caption[]{The deprojected thermodynamic profiles. In each panel, the subpanels show the deprojected density, the deprojected temperature, the deprojected pressure, the deprojected entropy, and the {\it projected} Fe abundance, which is obtained in the {\it projected} thermodynamic profiles and fixed to the best-fitting values in the spectral deprojection, from top to bottom. 0~arcsec in the horizontal axes corresponds to the location of the interface.}
 \label{img:prof_thermo_1}
\end{figure*}

\begin{figure*}
 \begin{minipage}{0.333\hsize}
  \centering
  \includegraphics[width=2.3in]{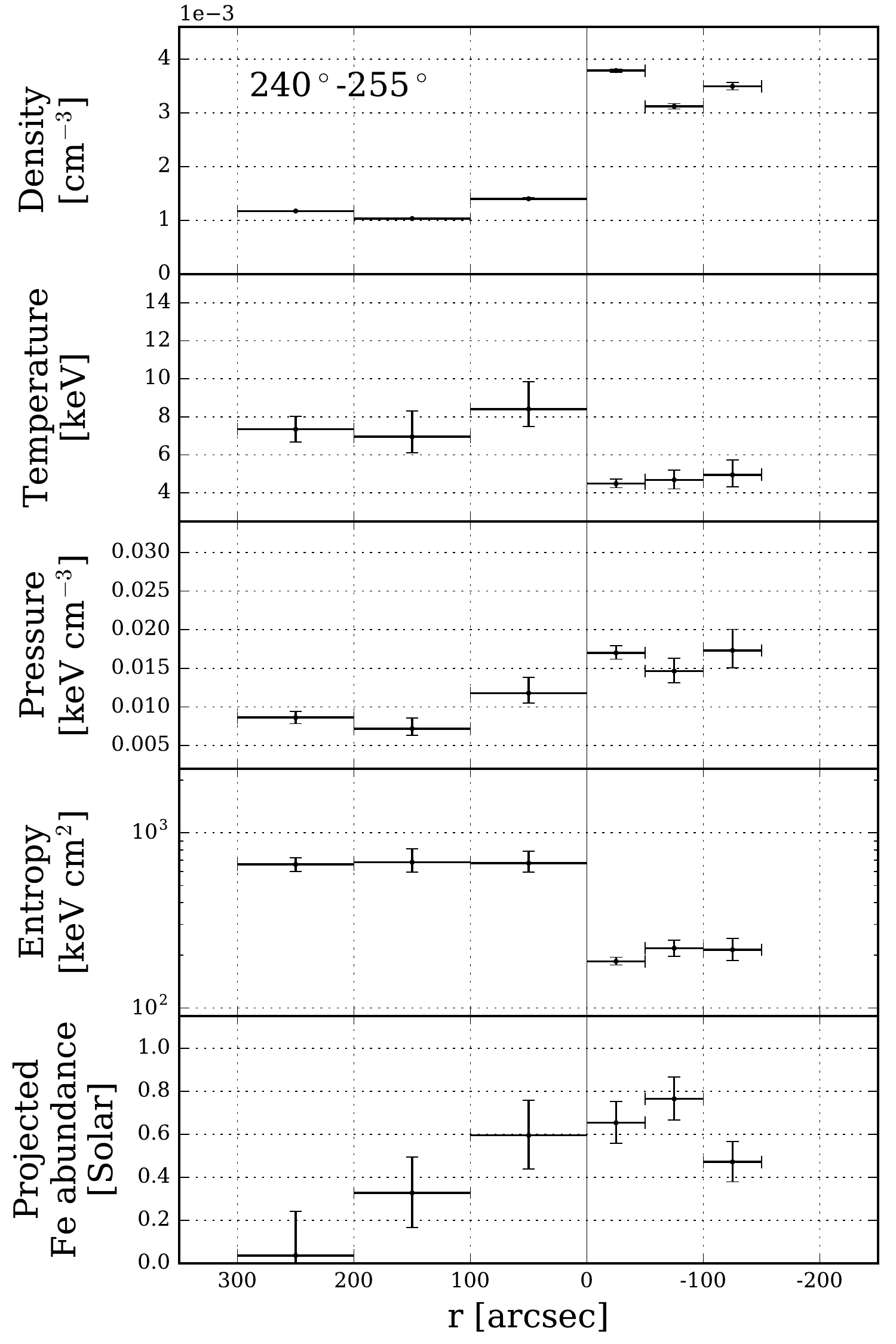}
 \end{minipage}%
 \begin{minipage}{0.333\hsize}
  \centering
  \includegraphics[width=2.3in]{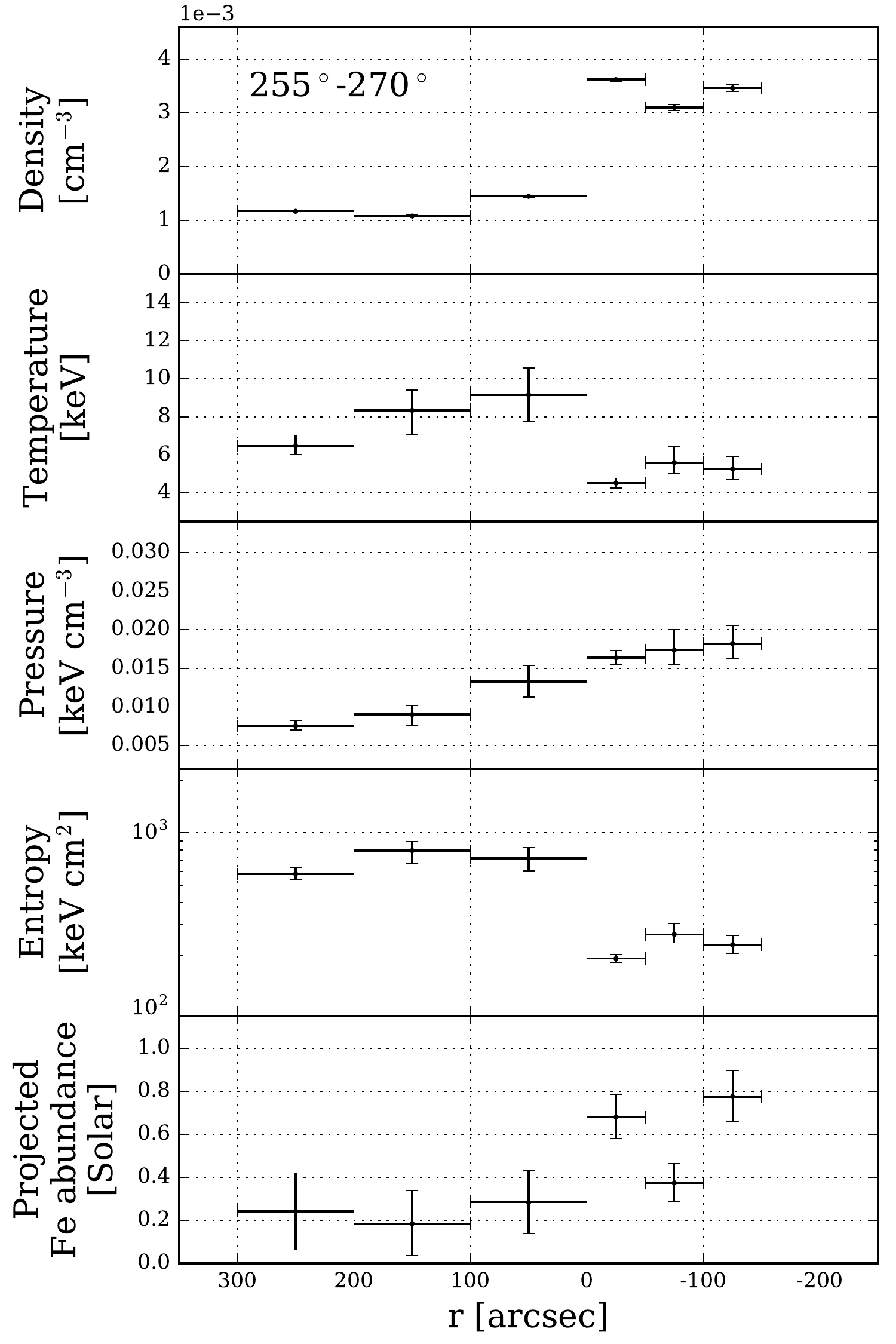}
 \end{minipage}%
 \begin{minipage}{0.333\hsize}
  \centering
  \includegraphics[width=2.3in]{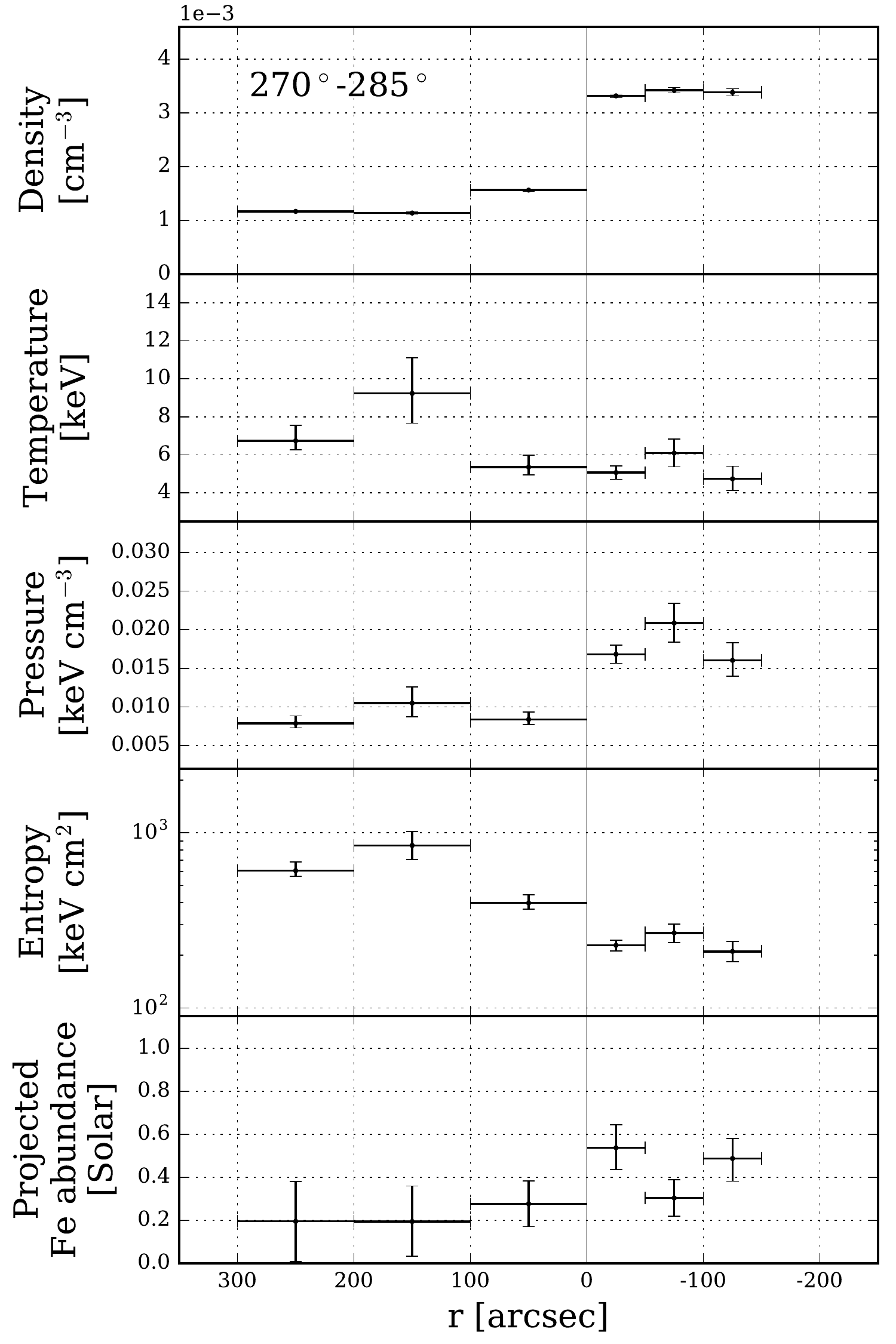}
 \end{minipage}
 \begin{minipage}{0.333\hsize}
  \centering
  \includegraphics[width=2.3in]{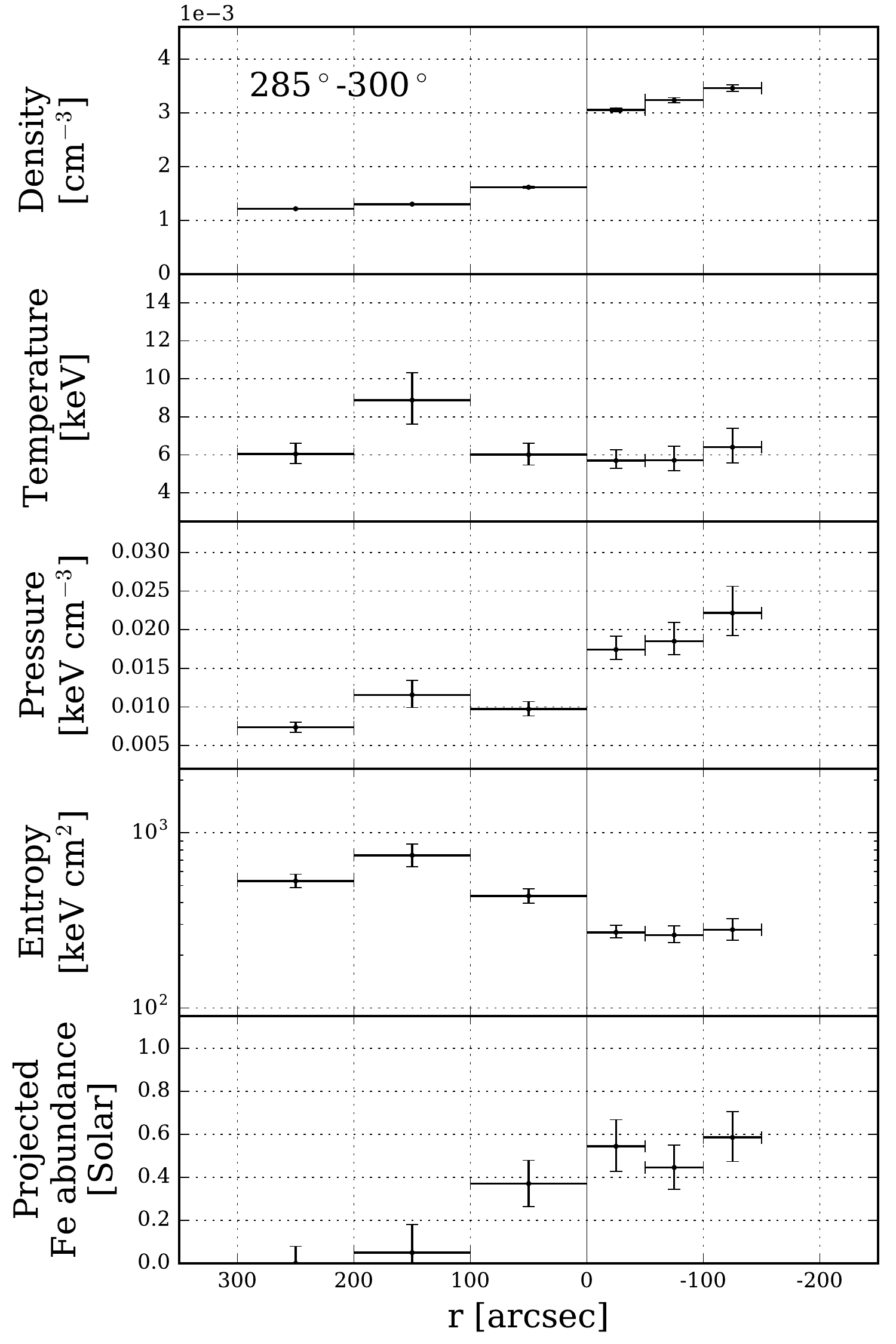}
 \end{minipage}
 \caption[]{Figure~\ref{img:prof_thermo_1}, continued.}
 \label{img:prof_thermo_2}
\end{figure*}



\bsp	
\label{lastpage}
\end{document}